\documentclass[a4paper,11pt]{article}

\pdfoutput=1 % if your are submitting a pdflatex (i.e. if you have
\usepackage{jheppub} % for details on the use of the package, please
\usepackage[T1]{fontenc} % if needed
\usepackage{braket}
\usepackage{xcolor}

\usepackage[en-US,useregional,showdow]{datetime2}
\DTMlangsetup[en-US]{ord=raise}

\usepackage{amsthm}
\theoremstyle{definition}

\theoremstyle{plain}

\usepackage[utf8]{inputenc}
\usepackage[T1]{fontenc}
\usepackage{graphicx}
\usepackage{caption}
\usepackage{subcaption}
\usepackage{grffile}
\usepackage{longtable}
\usepackage{wrapfig}
\usepackage{rotating}
\usepackage[normalem]{ulem}
\usepackage{amsmath}
\usepackage{textcomp}
\usepackage{amssymb}
\usepackage{capt-of}
\usepackage{hyperref}
\usepackage{amsthm}
\usepackage{color}
\usepackage{physics}
\usepackage{slashed}
\usepackage{siunitx}
\usepackage[capitalize]{cleveref}
\usepackage{slashed}

\usepackage{tikz}
\usetikzlibrary{calc,intersections,arrows.meta,
  decorations,
  decorations.pathmorphing,
  decorations.markings,
  shapes.geometric,
  hobby}

\usepackage{pgfplots}
\usepackage{pgfplotstable}
\usepgfplotslibrary{polar,fillbetween,groupplots}
\usepgfplotslibrary{external}
\newcommand{\ads}{\text{AdS}}
\newcommand{\AdS}{\text{AdS}}
\DeclareMathOperator{\sgn}{sign}
\DeclareMathOperator{\Pf}{Pf}
\newcommand{\pFq}[2]{{}_{#1}F_{#2}}

\newcommand{\autocite}[1]{\cite{#1}}

\title{\boldmath Size of bulk fermions in the SYK model}

\author[a]{Yuri D. Lensky,}
\author[a]{Xiao-Liang Qi}
\author[b]{and Pengfei Zhang}
\affiliation[a]{\small \em  Stanford Institute for Theoretical Physics, Stanford University, Stanford CA 94305 USA}
\affiliation[b]{\small \em California Institute of Technology,
Pasadena, CA 91125, U.S.A}
\emailAdd{ydl@stanford.edu}
\emailAdd{xlqi@stanford.edu}
\emailAdd{pengfeizhang.physics@gmail.com}

\abstract{%
  The study of quantum gravity in the form of the holographic duality has uncovered and motivated the detailed investigation of various diagnostics of quantum chaos. One such measure is the operator size distribution, which characterizes the size of the support region of an operator and its evolution under Heisenberg evolution. In this work, we examine the role of the operator size distribution in holographic duality for the Sachdev-Ye-Kitaev (SYK) model. Using an explicit construction of AdS$_2$ bulk fermion operators in a putative dual of the low temperature SYK model, we study the operator size distribution of the boundary and bulk fermions. Our result provides a direct derivation of the relationship between (effective) operator size of both the boundary and bulk fermions and bulk $\text{SL}(2; \mathbb{R})$ generators.
}

\begin{document}

\maketitle
\flushbottom
\tableofcontents

\section{Introduction}
\label{sec:introduction}

In recent years, significant progress has been made on characterizing quantum chaos in many-body systems. Developments in the study of holographic duality\cite{maldacena1999large} pointed out a close connection between chaotic many-body dynamics and gravitational physics, especially black hole dynamics\cite{sekino2008fast,shenker2014black}. Motivated by this connection, new characteristics of quantum chaos have been studied, such as the out-of-time-ordered correlator (OTOC)\cite{larkin1969quasiclassical,shenker2014black,shenker2014multiple,kitaev2014talk}. The OTOC can be viewed as a quantum generalization of the Poisson bracket in classical dynamics. In the classical case, the Poisson bracket of canonical coordinates $x(t)$ with the coordinates at an earlier time $x(0)$ determines how sensitive the trajectory is to initial conditions, which characterizes chaos and is related to Lyapunov exponents. Similarly, the OTOC provides a measure of ``operator scrambling'' -- how operators become more complicated under Heisenberg evolution. The decrease of the OTOC corresponds to an increase of the commutator between two operators $A(t)$ at time $t$ and $B(0)$ at time $0$, which captures that the support of $A(t)$ in operator space grows.

Another related measure of operator scrambling is the operator size distribution\cite{hosur2016characterizing,roberts2018operator}. By expanding each operator in a polynomial of simple operators, such as single Pauli operators in a spin chain, or single fermion creation/annihilation operators in a fermion system, one obtains a superposition of terms, each of which is a product of multiple simple building blocks. This leads to a definition of operator size distribution as the distribution of support over products of different lengths. A single Pauli operator in a spin chain has size $1$, while a product of two Pauli's on two sites has size $2$. An operator's size distribution provides a more sophisticated characteristic of its complexity than the OTOC. It was also shown that a particular average of OTOCs gives the average size, {\it i.e.} the first moment of the operator size distribution\cite{roberts2018operator,qi2019quantum}. The operator size distribution have been studied in various models including the Sachdev-Ye-Kitaev models\cite{roberts2018operator,qi2019quantum} and spin models\cite{qi2019measuring,xu2019locality,vermersch2019probing}. Finite temperature generalizations of (effective) operator size has been discussed recently\cite{lucas2019operator,mousatov2019operator}. Interestingly, the operator size distribution has also been related to a certain momentum quantum number in the holographic dual theory\cite{roberts2018operator,brown2018falling,Susskind:2018tei,Susskind:2019ddc,Lin:2019qwu}. 

Here, we investigate the role of operator size distribution in holographic duality by studying the bulk operator size in the dual theory of SYK model. The SYK model has been proposed to be dual to Jackiw-Teitelboim gravity\cite{jackiw1985lower,teitelboim1983gravitation} with certain bulk matter content. Although the duality is not completely proven, from the behavior of the boundary fermion in the SYK model (a conformal field with known conformal dimension, and large-$N$ factorization) it is reasonable to apply the known dictionary and determine the corresponding bulk fermion field. In this paper, we compute the size distribution of the boundary Majorana field in the low temperature, large $N$ limit of SYK, using a method inspired by a combination of \autocite{qi2019quantum} and \autocite{Gu:2017njx} that makes clear the connection of this boundary quantity to bulk quantities for a fermion in static $\ads_2$. Then, in close analogy with previous work for fields with other spin~\autocite{Hamilton:2005ju, Hamilton:2006az, Hamilton:2006fh, Kabat:2012hp} (often referred to collectively as the Hamilton-Kabat-Lifschytz-Lowe, or HKLL, constructions), we give an explicit construction of the bulk fermion operator in terms of the boundary Majoranas. This enables us to give a direct computation of the bulk fermion size, as well as a direct derivation of relationships between certain matrix elements of the boundary size operator, bulk fermion size, and certain components of bulk momentum ({\it i.e.} $\text{SL}(2,\mathbb{R})$ charge) in SYK models and their bulk duals. 

%This enables us to give a direct computation of the bulk fermion size, as well as provide a direct proof of conjectured relations between operator size and certain components of bulk momentum ({\it i.e.} $\text{SL}(2,\mathbb{R})$ charge) both in SYK models and their bulk duals.

% This allows us to compute the size distribution of the bulk fermion based on previous results about boundary fermion size distribution\cite{qi2019quantum}.
% We carried explicit calculation of the bulk fermion size, and our results provide an explicit proof of the proposed relation between operator size and a certain component of bulk momentum ({\it i.e.} $SL(2,R)$ charge). We also confirmed our results by numerical computation. 

The remainder of the paper is organized as follows. In Sec. \ref{sec:size-operator-distribution} we review the essential results on the definition of operator size distribution, and the generating function approach that will be useful for us. In Sec. \ref{sec:SYK} we review the key features of the SYK model, and derive the boundary operator size distribution in the ``low temperature'' limit ($N \gg \beta \mathcal{J} \gg 1$ in the SYK variables introduced in Section~\ref{sec:SYK}). In doing so, we directly find a relationship between the boundary size generating function at low temperature and $\text{SL}(2,\mathbb{R})$ generators. In Sec. \ref{sec:size-of-bulk-fields} we develop a construction of bulk fermions from the boundary, and use this to study the size of bulk fields in SYK. As a cross-check, we also present numerical results based on the known large-$q$ boundary size distribution~\autocite{qi2019quantum}. Finally, Sec. \ref{sec:conclusion} contains the conclusion and further discussion. Appendix~\ref{sec:syk-size-effective-action} gives details on the derivation of the boundary size. Appendix~\ref{sec:bulk-fermion-2pt} contains explicit expressions for expectations of momenta in static $\ads_2$. Appendices~\ref{sec:light-fermions} and \ref{sec:reconstruction-kernel-details} give the derivation of the bulk fermion reconstruction.

\section{The size operator and its distribution}
\label{sec:size-operator-distribution}
In this work, we will use the machinery developed in \cite{qi2019quantum} to treat operator size on the boundary. In this section, we present a brief overview of the setting and main results in that work.

For concreteness, imagine the space of operators in our quantum mechanical Hilbert space \(\mathcal{H}\) is generated by some finite collection of \(N\) Majorana operators, \(\{\chi_j, \chi_k\} = \delta_{jk}\). Any operator \(\mathcal{O}\) can be written
\begin{equation}
\label{eq:generic-op-chi-expansion}
\mathcal{O} = \sum_{n = 0}^{N} \sum_{1 \le j_1 < \cdots < j_n \le N} \mathcal{O}^{(n)}_{j_1 \cdots j_n} \chi_{j_1} \cdots \chi_{j_n}
\end{equation}
where the \(\mathcal{O}^{(n)}_{j_1 \cdots j_n}\) are complex numbers. We are interested in characterizing the ``weight'' of \(\mathcal{O}\) in different \(n\)-sectors. There is an abstract Hilbert space of operators, with Hilbert-Schmidt inner product \(\langle \mathcal{O}, \mathcal{S} \rangle = \Tr \mathcal{O}^{\dagger} \mathcal{S}\), and we choose to measure the size in some \(n\)-sector by the length of the projection of \(\mathcal{O}\) to the subspace of length-\(n\) \(\chi_j\) strings in this inner product.

It can be useful to work in the setting of a ``doubled Hilbert space'' \(\mathcal{H}^{(2)} = \mathcal{H} \otimes \mathcal{H}\), associated to two copies of the physical system, as opposed to the ``operator Hilbert space'', \(\mathcal{H}^{\text{op}} = \mathcal{H} \otimes \mathcal{H}^{*}\). This requires a (non-unique) choice of isomorphism between the two spaces. Fermionic or bosonic operators that appropriately commute or anti-commute with all operators on the right or left tensor factor of \(\mathcal{H}^{(2)}\) are written \(\mathcal{O}^L\) or \(\mathcal{O}^R\), respectively\footnote{Since we will need the fermionic statistics, we cannot simply use operators of the form \(\mathcal{O} \otimes 1\).}; if \(\{\chi_j\}\) generates the operator algebra on \(\mathcal{H}\), there are \(\{\chi^L_j\}\) that generate the same algebra (and satisfy the same relations as the \(\{\chi_j\}\) amongst themselves). In our case, we can take an irreducible representation of the Majorana algebra for \(2N\) fields (call this \(\mathcal{H}^{(2)}\)), and arbitrarily call \(N\) of them \(\chi^L_j\), and the other \(N\) \(\chi^R_j\). This gives a factorization of \(\mathcal{H}^{(2)} = \mathcal{H} \otimes \mathcal{H}\), where even products of \(\chi^L_j\) (\(\chi^R_j\)) act only on the left (right) factor. States in the doubled Hilbert space are written as \(| \psi )\). To implement our isomorphism, for an operator \(\mathcal{O} \in \mathcal{H}^{\text{op}}\), we define \(\mathcal{O}^L\) by expanding in terms of a generating set \(\{\chi_j\}\) and making the replacement \(\chi_j \to \chi^L_j\). Then, we choose a state \(| 0 ) \in \mathcal{H}^{(2)}\) and define \(| \mathcal{O} ) = \mathcal{O}^L | 0 )\). In order for this map to be injective, we require \(| 0 )\) to have full Schmidt rank between the two tensor factors, and to be proportional to an isometry, which requires the Schmidt weights must all be equal. In other words, \(| 0 )\) is a maximally entangled state between the two copies.

For our purpose, there is a particularly useful choice of maximally entangled state. Form the fermionic annihilation operators
\begin{equation*}
c_j = \frac{1}{\sqrt{2}}(\chi^L_j + i \chi^R_j)
\end{equation*}
and choose \(| 0 )\) such that \(c_j | 0 ) = 0\) for all \(j\).
It can be checked that this state is maximally entangled between the two tensor factors. Furthermore, this state has 
\begin{equation}
\label{eq:size-vacuum-chi-relations}
\chi^L_j | 0 ) = - i \chi^R_j | 0 ) = \frac{1}{\sqrt{2}} c_j^{\dagger} | 0 ).
\end{equation}
Then the information about the distribution of \(\mathcal{O}\) across operators of different size is contained in the moments of the numbers of \(\chi^L_j\) operators, or equivalently the \(c_j\) fermions,
\begin{align*}
n_j &= c_j^{\dagger} c_j = \frac{1}{2} + i \chi^L_j \chi^R_j, n = \sum_j n_j \\
n_j[\mathcal{O}]^{(k)} &= \frac{(\mathcal{O} | n_j^k | \mathcal{O})}{(\mathcal{O} | \mathcal{O})},
n[\mathcal{O}]^{(k)} = \frac{(\mathcal{O} | n^k | \mathcal{O})}{(\mathcal{O} | \mathcal{O})}.
\end{align*}
This is a state-independent measure of size, which does not allow the characterization of operator scrambling at a given energy scale or a subspace of states. To remedy this, Ref. \cite{qi2019quantum} proposed to measure the size in the thermal ensemble by moments of the generating function
\begin{equation}
Z_{\mu}^{\beta}[\mathcal{O}] = \frac{(\mathcal{O} \rho_{\beta}^{1/2}| e^{- \mu n} | \mathcal{O} \rho_{\beta}^{1/2})}
{(\rho_{\beta}^{1/2}| e^{- \mu n} | \rho_{\beta}^{1/2})}.\label{eq:sizepartitionfunc}
\end{equation}
The derivatives of the logarithm of this generating function over $\mu$ are the differences between the size cumulants for \(\mathcal{O} \rho_{\beta}^{1/2}\) and \(\rho_{\beta}^{1/2}\), for example the first moment is
\begin{equation}
n_{\beta}[\mathcal{O}] \equiv -\frac{\partial\log Z_\mu^\beta\left[\mathcal{O}\right]}{\partial\mu}= n[\mathcal{O} \rho_{\beta}^{1/2}] - n[\rho_{\beta}^{1/2}].\label{eq:effectivesize}
\end{equation}
The ``thermal size'' $n_\beta[\mathcal{O}]$ depends on the reference state $\rho_\beta$, which takes into account the fact that certain operators are more important than others when applying to a subspace of states.

Finally, we note that \(Z_{\mu}^{\beta}[\mathcal{O}(t)]\) is related to a particular ``thermal'' two-point function of the \(\mathcal{O}\). In particular, following Ref.\cite{qi2019quantum} we define the ``boundary size kernel''
\begin{equation}
\label{eq:boundary-size-kernel-def}
\mathcal{\tilde{G}}^{\partial}_{\mu}(\tau_1, \tau_2)
= \mathcal{G}^{\partial}_{\mu}(- i (\tau_1 - \frac{\beta}{2}), - i (\tau_2 - \frac{\beta}{2}))
= \frac{( 0 |\mathcal{T}[e^{- \left(\int_0^{\beta} d\tau H^L(\tau) + \delta(\tau - \beta/2) \mu n(\tau)\right)} \chi^L_j(\tau_1) \chi_j^L(\tau_2)]| 0 )}
{( 0 |\mathcal{T}[e^{- \left(\int_0^{\beta} d\tau H^L(\tau) + \delta(\tau - \beta/2) \mu n(\tau)\right)}]| 0 )},
\end{equation}
so that
\begin{equation*}
Z_{\mu}^{\beta}[\chi_j(t)] = \mathcal{G}^{\partial}_{\mu}(t - i \epsilon, t + i \epsilon).
\end{equation*}
The function \(\mathcal{G}^{\partial}_{\mu}\) can be computed as a single-sided quantity, as was discussed in Ref. \cite{qi2019quantum}.

\section{The Sachdev-Ye-Kitaev model}
\label{sec:SYK}
The Sachdev-Ye-Kitaev (SYK) model is an ensemble of Hamiltonians
\begin{equation}
\label{eq:syk-hamiltonian}
H_{SYK}[J] = i^{q/2} \sum_{1 \le j_1 < \cdots < j_q \le N} J_{j_1 \cdots j_q}
\chi_{j_1} \cdots \chi_{j_q},
\{\chi_j, \chi_k\} = \delta_{jk}
\end{equation}
where the \(J_{j_1 \cdots j_q}\) are independently drawn from normal distributions, with variance
\begin{equation*}
\langle J_{j_1 \cdots j_q}^2 \rangle
= \frac{(q-1)!}{N^{q-1}} J^2 
= \frac{2^{q-1} (q-1)!}{q N^{q-1}} \mathcal{J}^2.
\end{equation*}

This model is chaotic, in the sense that the out of time ordered four-point function grows exponentially, but is at the same time solvable in a \(1/N\) expansion. The SYK model has an emergent approximate reparametrization symmetry, which is explicitly broken by a UV cutoff term. At low temperature, the symmetry breaking is small, suppressed by $\frac1{\beta \mathcal{J}}$. The quasi-Goldstone modes of reparametrization symmetry breaking are governed by a Schwarzian action, which 
% . This symmetry is explicitly broken by the model \eqref{eq:syk-hamiltonian} itself in the UV, and is spontaneously broken in the IR. The leading action for the Goldstone modes associated to this symmetry breaking is the Schwarzian action. This 
is also the action (in appropriate variables) for Jackiw-Teitelboim (JT) two dimensional gravity with negative cosmological constant. In this sense, the SYK model is approximately dual to JT gravity. The complete bulk description is not known, but there is a possibility that the SYK model is an example of the \(\AdS\)/CFT duality between a \(d = 1\) ``nearly'' CFT and a ``nearly'' \(\AdS_2\) bulk described by JT gravity coupled to interacting matter fields~\autocite{Gross:2017hcz}.

Given the full boundary size kernel for the \(\chi_j\) fermions in the SYK model at low temperature, if we assume that there is a weakly coupled gravity dual to SYK, we can explicitly compute the size of the bulk fermions dual to the \(\chi_j\) operators. In fact, the boundary size kernel can be computed in two regimes; both at large \(q\), computed in \cite{qi2019quantum}, and at low temperature where the model is governed by the Schwarzian effective action. We discuss the size distribution in the low temperature, or Schwarzian, regime in the following section. In Section~\ref{sec:size-chi-fermions-generators} we describe a connection between the low temperature boundary size operator and bulk $\ads_2$ isometry generators that will be important for understanding the bulk size.

\subsection{Boundary operator size in SYK models}
\label{sec:lowt-syk-boundary-size}
In this section we examine the boundary size of the Majorana fermions in SYK in the low temperature limit, $N \gg \beta \mathcal{J} \gg 1$. To understand the bulk size, it is helpful to examine the derivation of the boundary size distribution in some detail. We work at large \(N\), so that in particular the non-local bulk interactions are suppressed. Then the SYK action can be written in terms of non-local fields \(G\) and \(\Sigma\). In particular, \(\Sigma\) is a Lagrange multiplier that enforces
\begin{equation*}
G(\tau_2, \tau_1) = \frac{1}{N} \sum_{j = 1}^N \chi_j(\tau_2) \chi_j(\tau_1),
\end{equation*}
where for convenience we take the arguments of \(G\) to be imaginary time.

As they are subleading in \(N\), we ignore the normal ordering constants in the effective action for the SYK model. We also assume that the model is totally self-averaging at leading order in \(N\), so that we can directly use the effective action after averaging over the couplings. Then the effective action for size (derived in more detail in Appendix \ref{sec:syk-size-effective-action}) is~\autocite{Kitaev:2017awl, Maldacena:2016hyu}
\begin{align}
S_s &= S_{\text{SYK}} + S_{\mu},
S_{\text{SYK}} = - \ln \Pf (\partial_{\tau} - \Sigma(\tau, \tau'))
+ \frac{1}{2} \int d\tau d\tau' \Sigma(\tau, \tau') G(\tau, \tau') - \frac{J^2}{q} G(\tau, \tau')^q \\
S_{\mu} &= \frac{\mu}{2} - \ln \cosh \frac{\mu}{2} - 2 \tanh \frac{\mu}{2} G\left(\frac{\beta}{2}, 0\right),\label{eq:Smu}
\end{align}
in the sense that the size distribution function can be computed as
\begin{equation}
G^{\partial}_{\mu}(\tau_2, \tau_1) = 
\frac{\int \mathcal{D} G \mathcal{D} \Sigma G(\tau_2, \tau_1) e^{- NS_s[G, \Sigma]}}
{\int \mathcal{D}G \mathcal{D} \Sigma e^{-NS_s[G, \Sigma]}}.
\end{equation}
At large \(\beta \mathcal{J}\), the low energy excitations of the SYK model can be thought of as reparametrizations of time \(\phi \to \theta(\phi)\), with Schwarzian action ~\autocite{Kitaev:2017awl, Maldacena:2016hyu, Bagrets:2016cdf}
\begin{equation*}
\tilde{S}_\text{SYK} = - \frac{2 \pi}{L} \int_0^{2 \pi} d \phi \{ \tan \frac{\theta(\phi)}{2}, \phi\},
L = \frac{\beta \mathcal{J}}{\alpha_S},
\end{equation*}
where \(\phi\) is related to the boundary time by \(\phi = 2 \pi \tau / \beta\), and $\alpha_S$ is a $q$-dependent constant computed in \autocite{Maldacena:2016hyu}.

Through this and the next section, we take a geometric approach to our computations, similar to the techniques of \autocite{Gu:2017njx}. It is useful to interpret the Schwarzian as the leading non-trivial part of the extrinsic curvature of a long curve in Euclidean \(\AdS_2\). In particular, in Rindler coordinates, for a curve \(\gamma(\phi) = (\theta(\phi), \rho(\phi))\) of large length \(L\) parameterized proportionally to arc length by an angular coordinate \(\phi \in [0, 2 \pi)\), the extrinsic curvature is
\begin{equation*}
K = 1 + \left( \frac{2 \pi}{L} \right)^2 \{ \tan \frac{\theta(\phi)}{2}, \phi \} + O(L^{-4}),
\end{equation*}
and by the Gauss-Bonnet theorem we find
\begin{equation*}
\int_{\partial M} K + \int_M \frac{R}{2} = 2 \pi
\implies S_{\text{SYK}} \approx L - A - 2 \pi
\end{equation*}
where $R = -2$ is the Ricci scalar, and \(A\) is the area of the region bounded by \(\gamma\). For this reason, the curve corresponding to the saddle point for \(\mu = 0\) is simply a circle with large length \(L\).

Under a reparametrization \(\phi \to \theta(\phi)\), the two-point function changes by
\begin{equation}
\label{eq:general-G-reparametrization}
G(\phi_2, \phi_1) \to [\theta'(\phi_2) \theta'(\phi_1)]^{\Delta} G(\theta(\phi_2), \theta(\phi_1)).
\end{equation}
When the \(\phi\) are sufficiently separated in imaginary time, we approximate the saddle point \(G\) by its conformal form,
\begin{align*}
G_c(\phi_2, \phi_1) &= c_{\Delta}
\left[ \left( \frac{2 \pi}{L} \right)^2 \frac{1}{2 \sin^2 \frac{\phi_2 - \phi_1}{2}} \right]^{\Delta} \sgn (\phi_2 - \phi_1) \\
b_{\Delta} &= \frac{1}{2} \left[ 2 \left( \frac{1}{2} - \Delta \right) \frac{\tan \pi \Delta}{\pi \Delta} \right]^{\Delta}, \; c_{\Delta} = \frac{b_{\Delta}}{(2 \alpha_S^2)^{\Delta}}
\end{align*}
and compute \(S_{\mu}\) on a reparametrization using \eqref{eq:general-G-reparametrization}. This shows that we can think of \(S_{\mu}\) as providing some ``tension'' between points on opposite sides of the circular saddle point solution, and for sufficiently small \(\mu\) it is self-consistent to compute around this saddle point. Because of the symmetry of the problem, the new saddle point will be approximated by a path which consists of two segments of equal
length, each of which is a portion of a circle \(C\) with the same fixed radius. Since we keep the length \(L\) fixed, we can parameterize the problem by a single number, the fraction of the circle \(C\) that makes up one of the two segments, in other words the angle \(\lambda\). We give an illustration of the saddle point in Figure~\ref{fig:SYK-size-saddle-point}. This solution is only an approximation: we consider small reparametrizations, which will keep the curve smooth, while the approximation has sharp corners at \(\phi = 0, \pi\).
\tikzset{op insertion/.style={inner sep=1pt, circle, fill=white}, ads boundary/.style={dotted, semithick, radius=1}}
\def\innerrad{.8}
\pgfmathsetmacro\circangle{pi + 1.3}
\pgfmathsetmacro\lenratalpha{(pi / \circangle) * (\innerrad / (1 - \innerrad * \innerrad))}
\pgfmathsetmacro\newrad{(-1 + sqrt(1 + 4 * \lenratalpha*\lenratalpha)) / (2 * \lenratalpha)}
\begin{figure}[thb]
  \centering
  \begin{subfigure}[b]{0.3\linewidth}
    \begin{tikzpicture}[scale=2]
      \draw[ads boundary] circle;
      \draw[radius=\innerrad] circle;
      \node[op insertion] (chi-0) at (0:\innerrad) {$\chi$};
      \node[op insertion] (chi-pi) at (180:\innerrad) {$\chi$};
      \draw[dashed,
      decoration={markings, mark=at position 0.25 with {\arrow{>}}, mark=at position 0.75 with {\arrow{<}}}, postaction=decorate] (chi-0) -- (chi-pi);
    \end{tikzpicture}
  \end{subfigure}
  \begin{subfigure}[b]{0.3\linewidth}
    \begin{tikzpicture}[scale=2]
      \draw[ads boundary] circle;
      \draw[radius=\innerrad, opacity=0.3] circle;
      \draw[radius=\newrad] ($({-(\circangle - pi)/2 * 180 / pi}:\newrad)$) coordinate (new-circ-start) arc[start angle = {-(\circangle - pi)/2 * 180 / pi}, delta angle = {\circangle * 180 / pi}] coordinate (new-circ-end);
      \draw[dashed] (0, 0) -- (new-circ-start);
      \draw[dashed] (0, 0) -- (new-circ-end);
      \draw[radius=.1] ($({-(\circangle - pi)/2 * 180 / pi}:0.1)$) arc[start angle = {-(\circangle - pi)/2 * 180 / pi}, delta angle = {\circangle * 180 / pi}];
      \node[above] at (0, 0.1) {$\lambda$};
    \end{tikzpicture}
  \end{subfigure}
  \begin{subfigure}[b]{0.3\linewidth}
    \begin{tikzpicture}[scale=2]
      \draw[ads boundary] circle;
      \draw[radius=\innerrad, opacity=0.3] circle;
      \draw plot[smooth] file {plot_data/saddle_solution_boundary_points.txt};
      \begin{scope}[yscale=-1]
        \draw plot[smooth] file {plot_data/saddle_solution_boundary_points.txt};
      \end{scope}
    \end{tikzpicture}
  \end{subfigure}
  \caption{Overview of the computation of the saddle point. We show Euclidean $\ads_2$ with radial coordinate $r = \tanh \frac{\rho}{2} \in [0, 1)$, with the conformal boundary drawn as a dotted line. In the first panel, we draw the saddle point at $\mu = 0$, which is a circle of length $L$. We also show the operator insertions at $\phi = 0, \pi$ that give ``tension'' to the solution in the $\mu \ne 0$ case. In the presence of these operator insertions, the true saddle is a shape that is ``pinched'' towards the center. The solution will consist of two circular segments. In the second panel, we show the top circular segment in a coordinate where its center is at $r = 0$, with the $\mu = 0$ saddle for reference. The length of each segment is fixed to $L / 2$, so for an inner angle $\lambda > \pi$, the radius of the circle must shrink accordingly. In the last panel, we show the $\mu \ne 0$ saddle in a coordinate that is symmetric between the two segments, namely one where $\phi = 0, \pi$ coincide with $\theta = 0, \pi$ and these points are equidistant from $r = 0$, where $\phi \in [0, 2 \pi)$ parameterized the saddle curve proportionally to arc length. The transformation of coordinates between the first and second panel is a boost in embedding coordinates that moves the endpoints of the segment to the $\theta = 0, \pi$ line, given explicitly in \eqref{eq:upper-centered-to-symmetric-boost}.}
  \label{fig:SYK-size-saddle-point}
\end{figure}
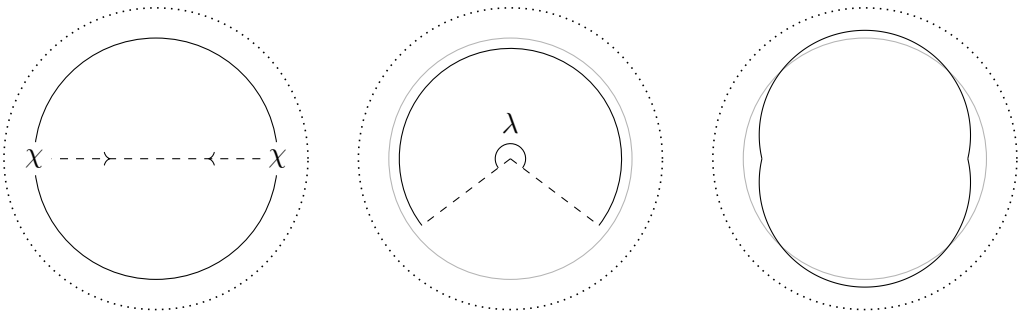

We reiterate that on the saddle point solution, times in the boundary theory are related to points in Euclidean $\ads_2$ by the point on the saddle point curve at parameter $\phi = 2 \pi \tau / \beta$. To compute the two-point function on this saddle, we use the relation \eqref{eq:g-as-h2-distance}, which gives (for $\tau_2$ later on the thermal circle than $\tau_1$)
\begin{equation}
  G_{\mu}^{\partial}(\tau_2, \tau_1) \approx c_{\Delta} (\cosh \delta_{21} - 1)^{- \Delta},
\end{equation}
where $\delta$ is the geodesic distance between the points corresponding to $\tau_2$ and $\tau_1$ on the saddle point curve. For the sizes of the thermal state and boundary fermion, it remains to find the dependence of the angle \(\lambda\) on \(\mu\). In Appendix \ref{sec:syk-size-effective-action} we find this to be
\begin{equation*}
\delta \lambda = 2 \tanh \frac{\mu}{2} \frac{\Delta c_{\Delta}}{2^{\Delta}} \left( \frac{L}{2\pi} \right)^{1 - 2 \Delta}
- \left( 2 \tanh \frac{\mu}{2} \right)^2 \frac{1 - 2 \Delta}{\pi} \left( \frac{\Delta c_{\Delta}}{2^{\Delta}} \right)^2
\left( \frac{L}{2\pi} \right)^{2 - 4 \Delta}
+ O((2 \tanh \frac{\mu}{2} L^{1 - 2 \Delta})^{3}).
\end{equation*}
Our main interest is in the first moment of size, for which we only need the first term in this expression.
In general, for small \(\mu\), \(\delta \lambda\) can be expanded order-by-order in powers of \(2 \tanh \frac{\mu}{2} L^{1 - 2 \Delta}\) (the full expression can be found in Appendix~\ref{sec:syk-size-effective-action}). We note $\delta \lambda$ is small for small $\mu < \alpha_S / \beta \mathcal{J}$, so in this regime our expansion around the $\mu = 0$ saddle point is self-consistent.

The cumulants of the size distribution of the thermal state \(\rho_{\beta}^{1/2}\) are the derivatives of the action \(N S_{\mu}\). For example, the average size is
\begin{equation*}
\frac{n_{\beta}}{N} = \frac{1}{2} - b_{\Delta} \left( \frac{\pi}{\beta \mathcal{J}} \right)^{2 \Delta}.
\end{equation*}
We can see that the \(n\)th cumulant will be of order \(N\), but is in general of order \((\beta \mathcal{J})^{n(1-2\Delta) - 1}\) in coupling. Since we take \(\beta \mathcal{J} \ll N\), we can consider the fluctuations in size of the thermal state to be suppressed by \((\beta \mathcal{J})^{\frac{1}{2} - 2 \Delta} / \sqrt{N}\). We note that this matches the results in the large-\(q\) limit\cite{qi2019quantum}.

\subsection{Size of $\chi$ fermions and $\text{SL}(2,\mathbb{R})$ generators}
\label{sec:size-chi-fermions-generators}

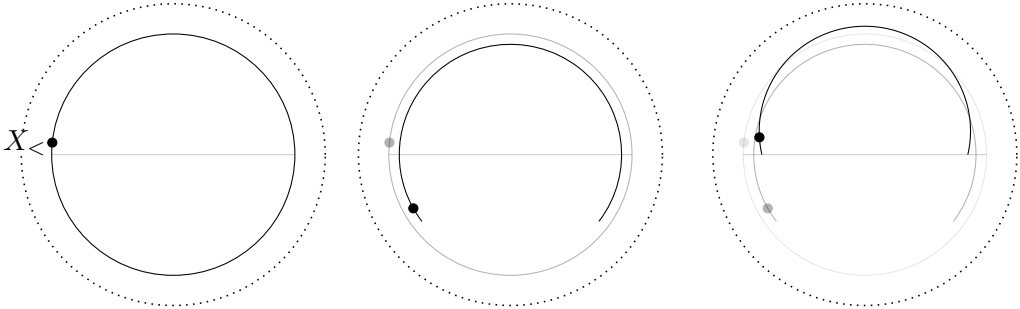
\begin{figure}[thb]
  \centering
  \def\imsplit{0.1}
  \def\xminusthetafinal{2.9774103916083163}
  \def\xminusrfinal{0.7032827329031786}
  \tikzset{xminuspoint/.style={radius=0.03,fill=black},
  thetazeroline/.style={opacity=0.2}}
  \begin{subfigure}[b]{0.3\linewidth}
    \begin{tikzpicture}[scale=2]
      \draw[ads boundary] circle;
      \draw[radius=\innerrad] circle;
      \draw[xminuspoint] ($({(pi - \imsplit) * 180 / pi}:\innerrad)$) circle node[shape=coordinate] (initial-xminus) {};
      \node[left=2pt, fill=white, inner sep=0pt] at (initial-xminus) {$X_{<}$};
      \draw[thetazeroline] (\innerrad, 0) -- (180:\innerrad);
    \end{tikzpicture}
  \end{subfigure}
  \begin{subfigure}[b]{0.3\linewidth}
    \begin{tikzpicture}[scale=2]
      \draw[ads boundary] circle;
      \begin{scope}[opacity=0.3]
        \draw[radius=\innerrad] circle;
        \draw[xminuspoint] ($({(pi - \imsplit) * 180 / pi}:\innerrad)$) circle node[shape=coordinate] (initial-xminus) {};
      \end{scope}
      \draw[radius=\newrad] ($({-(\circangle - pi)/2 * 180 / pi}:\newrad)$) arc[start angle = {-(\circangle - pi)/2 * 180 / pi}, delta angle = {\circangle * 180 / pi}];
      \draw[xminuspoint] ($({(\circangle / pi * (pi - \imsplit) - (\circangle - pi) / 2) * 180 / pi}:\newrad)$) circle;
      \draw[thetazeroline] (\innerrad, 0) -- (180:\innerrad);
    \end{tikzpicture}
  \end{subfigure}
  \begin{subfigure}[b]{0.3\linewidth}
    \begin{tikzpicture}[scale=2]
      \draw[ads boundary] circle;
      \begin{scope}[opacity=0.1]
        \draw[radius=\innerrad] circle;
        \draw[xminuspoint] ($({(pi - \imsplit) * 180 / pi}:\innerrad)$) circle node[shape=coordinate] (initial-xminus) {};
      \end{scope}
      \begin{scope}[opacity=0.3]
        \draw[radius=\newrad] ($({-(\circangle - pi)/2 * 180 / pi}:\newrad)$) arc[start angle = {-(\circangle - pi)/2 * 180 / pi}, delta angle = {\circangle * 180 / pi}];
        \draw[xminuspoint] ($({(\circangle / pi * (pi - \imsplit) - (\circangle - pi) / 2) * 180 / pi}:\newrad)$) circle;
      \end{scope}
      \draw plot[smooth] file {plot_data/saddle_solution_boundary_points.txt};
      \draw[xminuspoint] ($(\xminusthetafinal * 180 / pi:\xminusrfinal)$) circle;
      \draw[thetazeroline] (\innerrad, 0) -- (180:\innerrad);
    \end{tikzpicture}
  \end{subfigure}
  \caption{Illustration of the geometrical approach to the first size moment computation. In the first panel, we show the $\mu = 0$ location of the point $X_{<}$, which will lie on the first segment of the $\mu > 0$ solution. In this coordinate, it lies at some $\theta_0 < \pi$. We indicate the location of the point corresponding to $X_{<}$ with a dark dot, and show its previous location in a lighter color. Throughout, we show only the top segment of the $\mu \ne 0$ saddle. A line of points at $\theta = 0, \pi$ is drawn for reference. In the second panel, we use a coordinate so that the segment $X_{<}$ lies on is centered at $r = 0$. The boundary time is an affine parameter for the saddle point solution, so $X_{<}$ now lies at angle $\theta_1 = \frac{\lambda}{\pi} \theta_0$. In the last panel, we have changed to the more symmetric coordinate of the third panel of Figure \ref{fig:SYK-size-saddle-point} by the boost \eqref{eq:upper-centered-to-symmetric-boost}. Note that we can approximate the first move of $X_{<}$ by a rotation, generated by $B^{(E)}$, and the second is the coordinate transformation boost, generated by $E^{(E)}$. A point $X_{>}$ on the second segment, with some $\theta_0 > \pi$ on the $\mu = 0$, is transformed similarly (of course with the opposite boost). Once we compute the positions of the points in the symmetric coordinate system, we can transform by a final isometry to restore the position of one of the points, say $X_{>}$.}
  \label{fig:first-size-moment-geometry}
\end{figure}

The geometrical picture allows us to not only compute the size of boundary fermions, but also to uncover directly the relation of the size operator to bulk isometry generators. A similar relationship for the ``diagonal'' matrix elements of the size operator, $n_\beta[\chi(u)]$, was found by~\autocite{Lin:2019qwu}. For the discussion of bulk size, we will need the more general matrix elements $(\chi(u) \rho_{\beta}^{1/2} | n | \chi(u') \rho_{\beta}^{1/2})$, and in finding their relationship to isometry generators we also give another direct derivation of the result in~\autocite{Lin:2019qwu}. The key idea is that matrix elements of the size operator are determined by the change of the two-point function $G_{\mu}^{\partial}$ as a function of $\mu$. The two-point function on the saddle is approximately a function of the geodesic distance (as in \eqref{eq:g-as-h2-distance}), so $\mu$ affects the two-point function by deforming the boundary curve in Figure~\ref{fig:SYK-size-saddle-point}, which changes the distance between these two points in the bulk. Therefore the $\mu$ dependence can be mapped to a relative motion of the two points geometrically, which can be achieved by applying the bulk isometry transformations to one of the two points while keeping the other point fixed~\footnote{The self-consistency of this approximation is argued as follows. In Section~\ref{sec:lowt-syk-boundary-size} and Appendix~\ref{sec:syk-size-effective-action}, we show that for small enough $\mu$ it is self-consistent to compute the saddle point of $S_s$ about the $\mu = 0$ solution, and we assume that $\mu$ is in this regime for this section as well. On this saddle point, $G$ measures the geodesic distance between points. As long as we do not scale our point splitting with $N$, at leading order in $N$ there will not be an additional modification to the saddle point resulting from insertion of $G$ into the path integral.}. The key elements of the computation in this section are illustrated in
Figure~\ref{fig:first-size-moment-geometry}.

Euclidean $\ads_2$ (the hyperbolic plane) can be isometrically embedded in $\mathbb{R}^{(1,2)}$ ($\mathbb{R}^3$ with metric $(-1, 1, 1)$ and coordinate labels $(X^{(0)}, X^{(1)}, X^{(2)})$) as the surface $X^2 = -1$. Likewise, Lorentzian $\ads_2$ is the surface $X^2 = -1$ in $\mathbb{R}^{(2, 1)}$ ($\mathbb{R}^3$ with metric $(-1, -1, 1)$ and coordinate labels $(X^{(L0)}, X^{(L1)}, X^{(L2)})$). Unless otherwise noted, we will always use these coordinates (in this ordering) on the embedding space. In these coordinates, the matrices generating the independent isometries on Euclidean $\ads_2$ are
\begin{eqnarray}
E^{(E)} =
\begin{pmatrix}
0 & 1 & 0 \\
1 & 0 & 0 \\
0 & 0 & 0
\end{pmatrix},
B^{(E)} =
\begin{pmatrix}
0 & 0 & 0 \\
0 & 0 & 1 \\
0 & -1 & 0
\end{pmatrix},
P^{(E)} =
\begin{pmatrix}
0 & 0 & 1 \\
0 & 0 & 0 \\
1 & 0 & 0
\end{pmatrix},\label{eq:generators}
\end{eqnarray}
whose flows are shown in Figure~\ref{fig:euclidean-generators}.
\tikzset{vflow arrow/.style={-stealth, very thick}}
\begin{figure}[thb]
  \centering
  \def\plotwidth{2.5in}
  \pgfplotsset{every axis/.append style={axis lines=none,
      width=\plotwidth, height=\plotwidth,
      samples=17, samples y=5,
      domain=0:360,
      domain y=.27:.9,
      ymax=1.1},
    every axis plot/.style={vflow arrow, quiver={scale arrows = 0.2}}}
  \begin{subfigure}[b]{0.3\linewidth}
    \begin{tikzpicture}
      \begin{polaraxis}
        \addplot3[
        quiver={
          u={y * 180 / pi},
          v={0},
        }] {0};
        \draw[ads boundary] let \p1=($(axis cs:0, 1) - (axis cs:0, 0)$)
        in (axis cs:0, 0) circle [radius=\x1];
      \end{polaraxis}
    \end{tikzpicture}
  \end{subfigure}
  \begin{subfigure}[b]{0.3\linewidth}
    \begin{tikzpicture}
      \begin{polaraxis}
        \addplot3[
        quiver={
          u={(1 + y * y) * cos(x) * 180 / pi},
          v={sin(x) * (1 - y * y)},
          scale arrows = 0.2,
        }] {0};
        \draw[ads boundary] let \p1=($(axis cs:0, 1) - (axis cs:0, 0)$)
        in (axis cs:0, 0) circle [radius=\x1];
      \end{polaraxis}
    \end{tikzpicture}
  \end{subfigure}
  \begin{subfigure}[b]{0.3\linewidth}
    \begin{tikzpicture}
      \begin{polaraxis}
        \addplot3[
        quiver={
          u={- (1 + y * y) * sin(x) * 180 / pi},
          v={cos(x) * (1 - y * y)},
        }] {0};
        \draw[ads boundary] let \p1=($(axis cs:0, 1) - (axis cs:0, 0)$)
        in (axis cs:0, 0) circle [radius=\x1];
      \end{polaraxis}
    \end{tikzpicture}
  \end{subfigure}
  \caption{Vector flows of the Euclidean symmetry generators $B^{(E)}$, $E^{(E)}$, and $P^{(E)}$.}
  \label{fig:euclidean-generators}
\end{figure}
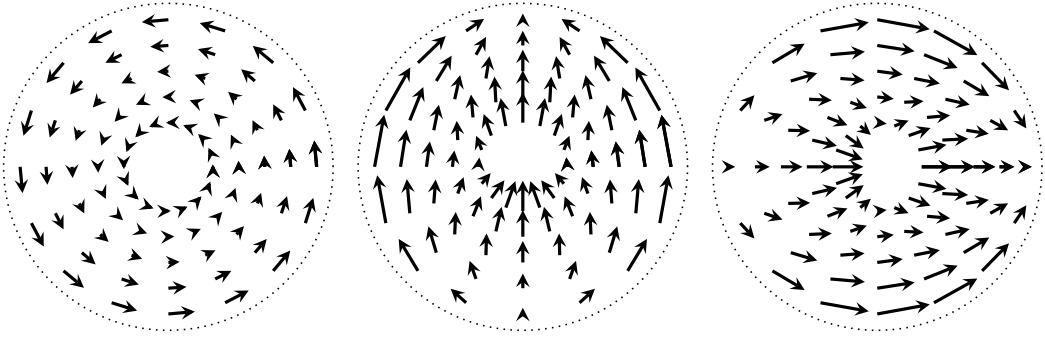
A symmetric coordinate system to consider our problem is one where the two distinct segments meet at \(\theta = 0\) and \(\pi\). We must consider the motion of two points, \(X_{<}\) on the first arc, and \(X_{>}\) on the second, as we perturb \(\lambda\). This is easiest to compute by placing the center of the circle our point is on at \(\rho = 0\), computing the location of \(X_{\gtrless}\) in that coordinate for the given \(\lambda\), then performing a boost by \(E^{(E)}\) to move the segment to its final position. The result is that (see Figure~\ref{fig:first-size-moment-geometry} and Appendix~\ref{sec:syk-size-effective-action} for more details)
\begin{multline}
\label{eq:xpm-difference-direct-derivative}
\frac{d}{d\lambda} (X_{>} - X_{<})
= - (\tanh \rho E^{(E)} + B^{(E)}) \left( \frac{X_{>} + X_{<}}{2} \right)
\\
- \frac{\tanh \rho}{\pi} \left[ E^{(E)}(\sin \phi_+ X_{>} - \sin \phi_- X_{<})
+ P^{(E)} (\cos \phi_+ X_{>} - \cos \phi_- X_{<})
 \right] \\
+ \frac{B^{(E)}}{\pi} [(\phi_+ - \pi) X_{>} - (\phi_- - \pi) X_{<}].
\end{multline}
Alternatively, we could have considered a coordinate system where \(X_{>}\) remains fixed as a function of \(\lambda\); this amounts to the replacement \(X_{>} \to X_{<}\) on the right side of \eqref{eq:xpm-difference-direct-derivative}. Analytically continuing the derivative to Lorentzian signature and using \(\phi_- = \pi - \epsilon + i t_1\), \(\phi_+ = \pi + \epsilon + i t_2\), we find
\begin{multline}
\label{eq:xminus-lambda-derivative}
\left. \frac{d}{d\lambda} X_{<}^{(L)} \right|_{\lambda = \pi}
=
\left[ i \left(\tanh \rho E^{(L)} - \left(1 + \frac{2 \epsilon}{\pi}\right) B^{(L)} \right)
\right. \\ \left.
+ \frac{\tanh \rho}{\pi} \left( (\sinh (t_2 - i \epsilon) - \sinh (t_1 + i \epsilon)) E^{(L)}
- (\cosh (t_2 - i \epsilon) - \cosh (t_1 + i \epsilon))P^{(L)} \right) 
\right. \\ \left.
- \frac{t_2 - t_1}{\pi} B^{(L)}
\right] X_{<}^{(L)}.
\end{multline}
where the infinitesimal Lorentzian generators in our standard coordinates are
\begin{equation*}
E^{(L)} =
\begin{pmatrix}
0 & -1 & 0 \\
1 & 0 & 0 \\
0 & 0 & 0
\end{pmatrix},
B^{(L)} =
\begin{pmatrix}
0 & 0 & 0 \\
0 & 0 & 1 \\
0 & 1 & 0
\end{pmatrix},
P^{(L)} =
\begin{pmatrix}
0 & 0 & 1 \\
0 & 0 & 0 \\
1 & 0 & 0
\end{pmatrix}.
\end{equation*}
Flows of these generators are shown in Figure~\ref{fig:lorentzian-vector-flows}.
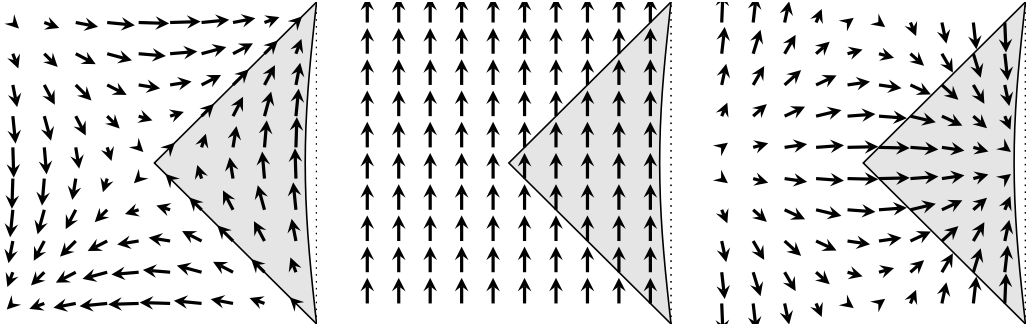
\begin{figure}[thb]
  \centering
  \def\buffer{.2}
  \def\plotwidth{2.3in}
  \def\boundrho{2.9}
  \pgfmathsetmacro\coshboundrho{0.5 * (e^(\boundrho) + e^(-\boundrho))}
  \tikzset{wedge outline/.style={-, semithick}}
  \pgfplotsset{
    every axis/.append style={view={0}{90}, axis lines=none,
      width=\plotwidth, height=\plotwidth,
      min = -pi/2, max=pi/2,
      domain = -(pi/2 - \buffer):(pi/2 - \buffer),
      domain y = -(pi/2 - \buffer):(pi/2 - \buffer),
      samples = 10,
      samples y = 10},
    wedge outline plot/.style={wedge outline,
      smooth, domain=-pi/2:pi/2},
    every axis plot/.append style={
      vflow arrow,
      quiver={scale arrows=0.305}}
  }
  \tikzset{rindler wedge/.style={draw=none, fill=gray, opacity=0.2}}
  \begin{subfigure}[b]{0.3\linewidth}
    \begin{tikzpicture}
      \begin{axis}
        \draw [ads boundary] (axis cs:pi/2,-pi/2) -- (axis cs:pi/2,pi/2);
        \addplot[wedge outline plot, name path=rightboundary]
        ({pi / 180 * acos(cos(x * 180 / pi) / \coshboundrho)}, x);
        \draw [wedge outline, name path=leftboundary] (axis cs:pi/2,-pi/2) -- (axis cs:0, 0) -- (axis cs:pi/2,pi/2);
        \addplot [rindler wedge] fill between [of=leftboundary and rightboundary];

        \addplot3 [
        quiver={
          u={cos(x * 180 / pi) * sin(y * 180 / pi)},
          v={cos(y * 180 / pi) * sin(x * 180 / pi)},
        }
        ] {0};
      \end{axis}
    \end{tikzpicture}
  \end{subfigure}
  \begin{subfigure}[b]{0.3\linewidth}
    \begin{tikzpicture}
      \begin{axis}
        \draw [ads boundary] (axis cs:pi/2,-pi/2) -- (axis cs:pi/2,pi/2);
        \addplot[wedge outline plot, name path=rightboundary]
        ({pi / 180 * acos(cos(x * 180 / pi) / \coshboundrho)}, x);
        \draw [wedge outline, name path=leftboundary] (axis cs:pi/2,-pi/2) -- (axis cs:0, 0) -- (axis cs:pi/2,pi/2);
        \addplot [rindler wedge] fill between [of=leftboundary and rightboundary];

        \addplot3 [
        quiver={
          u={0},
          v={1},
          scale arrows=0.25
        }
        ] {0};
      \end{axis}
    \end{tikzpicture}
  \end{subfigure}
  \begin{subfigure}[b]{0.3\linewidth}
    \begin{tikzpicture}
      \begin{axis}
        \draw [ads boundary] (axis cs:pi/2,-pi/2) -- (axis cs:pi/2,pi/2);
        \addplot[wedge outline plot, name path=rightboundary]
        ({pi / 180 * acos(cos(x * 180 / pi) / \coshboundrho)}, x);
        \draw [wedge outline, name path=leftboundary] (axis cs:pi/2,-pi/2) -- (axis cs:0, 0) -- (axis cs:pi/2,pi/2);
        \addplot [rindler wedge] fill between [of=leftboundary and rightboundary];

        \addplot3 [
        quiver={
          u={cos(x * 180 / pi) * cos(y * 180 / pi)},
          v={- sin(y * 180 / pi) * sin(x * 180 / pi)}
        }
        ] {0};
      \end{axis}
    \end{tikzpicture}
  \end{subfigure}
  \caption{Vector flows of the Lorentzian generators $B^{(L)}$, $E^{(L)}$, and $P^{(L)}$ on a patch of global $\ads_2$. For reference, we have drawn in a portion of the Rindler wedge with finite boundary location, as well as the right conformal boundary.}
  \label{fig:lorentzian-vector-flows}
\end{figure}

We can now compute the size in the low temperature limit. For convenience, we start with the Euclidean expression (keeping only the leading order in \(L \sim \beta \mathcal{J}\))
\begin{align}
- \frac{d}{d\mu} \ln G^{\partial}_{\mu}(\pi + \epsilon + i \phi, \pi - \epsilon + i \phi) 
&= \frac{\Delta^2 c_{\Delta}}{2^{\Delta}} \left( \frac{L}{2 \pi} \right)^{1 - 2 \Delta} \frac{1}{\frac{1}{2} (X_{>} - X_{<})^2} (X_{>} - X_{<}) \cdot \frac{d}{d \lambda} (X_{>} - X_{<})\nonumber\\
&= - \frac{\Delta^2 c_{\Delta}}{2^{\Delta}} \left( \frac{L}{2 \pi} \right)^{1 - 2 \Delta} \frac{1}{\frac{1}{2} (X_{>} - X_{<})^2} (X_{>} - X_{<}) \cdot (E^{(E)} + B^{(E)}) X_{<} \nonumber\\
&= \frac{\Delta^2 c_{\Delta}}{2^{\Delta}} \left( \frac{L}{2 \pi} \right)^{1 - 2 \Delta} 
\frac{\cosh \phi - \cos \epsilon}{\sin \epsilon}.
\end{align}
As noted in \cite{qi2019quantum}, the size at \(\phi = 0\) should be given by \(2 G_{0}^{\partial}(\pi, 0)\). We can use this to fix a UV regulator so that the size units match at \(\phi = 0\), and find \(\epsilon = \frac{8 \pi}{\Delta^2 L}\). Thus we find the boundary size at low temperature is
\begin{equation}
\label{eq:boundary-size-behaviour}
n_{\beta}[\chi(u)] = 2 b_{\Delta} \left( \frac{\pi}{\beta \mathcal{J}} \right)^{2 \Delta}
\left( 1 + \left( \frac{\Delta^2}{\alpha_S} \right)^2 \left( \frac{\beta \mathcal{J}}{4 \pi} \right)^2 \sinh^2 \left( \frac{\pi u}{\beta} \right) \right)
\end{equation}
where we have written the size in boundary time units \(u\). In general, in terms of the behaviour of size, the regulator simply sets the units as long as it is \(\epsilon \sim \pi / \beta \mathcal{J}\). We took the case \(u_1 = u_2\), and so were able to ignore contributions that vanish in this limit. Evaluating the derivative for \(u_1 \ne u_2\) gives the general Schwarzian contribution to the four-point function, which was also pointed out to be given by applying symmetry generators to one time argument of the conformal two-point function in~\cite{kitaev2018soft}. We note that the result \eqref{eq:boundary-size-behaviour} matches the low-temperature limit of \cite{qi2019quantum}.

We mention that this method extends to higher moments of size. These moments depend on higher order derivatives of $G_{\mu}^{\partial}$ over $\lambda$. From the exact bulk location of $X_{\gtrless}$ as a function of $\lambda$ (computed in Appendix~\ref{sec:syk-size-effective-action}), we just compute the derivatives of these points over $\lambda$ to the required order. We then compute the derivatives of $\lambda$ over $\mu$ to the same order, and differentiate $G_{\mu}^{\partial}$ to find the size moment to the desired order.

In addition to providing an analytic computation of operator size distribution in the low temperature region, the discussion in this subsection also gives a direct relation of boundary operator size with $\text{SL}(2,\mathbb{R})$ generators. If we consider a bulk dual fermion, the correlation function of which reproduces that of the boundary fermion for points approaching the boundary, we can also relate the boundary operator size to an $\text{SL}(2,\mathbb{R})$ quantum number of the bulk fermion. This is a warm-up calculation for the bulk operator size results in the next section, but we present it here since it does not depend on any bulk reconstruction, and is closely related to the previous part of this subsection.

Since the two-point function of a bulk fermion in static \(\ads_2\) approaches \(G_c\) as \(\rho \to \infty\) (c.f. Appendix \ref{sec:bulk-fermion-2pt}), from \eqref{eq:xminus-lambda-derivative} we can conclude that the boundary size at low temperature is given by the bulk expectation of generators for a free fermion, taken to the boundary and normalized by the appropriate factors. Explicitly, define
\begin{align}
\label{eq:size-matrix-element-generators}
S &= E - B \\
J_1 &= - \frac{i}{\pi} ((\sinh(t_2 - i \epsilon) - \sinh(t_1 + i \epsilon)) E - (\cosh(t_2 - i \epsilon) - \cosh(t_1 + i \epsilon)) P) \\
J_2 &= i \frac{t_2 - t_1}{\pi} B.
\end{align}
If we take the natural Rindler vielbien, there is a particular fermion component, say with index \(j\), whose two-point function vanishes slower as \(\rho \to \infty\) (corresponding to the eigenvalue of \(\gamma^1\) that does not vanish in the limit of \eqref{eq:large-rho-bulk-correlator-limit}; if we take the bulk mass positive (negative) this is the \(+1\) (\(-1\)) eigenvector). Then we find, writing \(\langle \cdot \rangle\) to mean expectation values for a free fermion in the Poincare vacuum on \(\ads_2\),
\begin{align}
\cosh \rho_0 &\gg \frac{L}{2\pi}, \, x_j = (t_j, \rho_0), \,
B_{\Delta} = \frac{b_{\Delta}}{N_{\Delta, 1} (1 - 2 \Delta)} \left( \frac{\pi}{\beta \mathcal{J}} \right)^{2 \Delta} \\
\label{eq:boundary-size-bulk-generators}
- \left. \frac{d}{d\mu} \right|_{\mu = 0} G_{\mu}^{\partial}(t_2, t_1) &= 
(\chi(t_2) \rho_{\beta}^{1/2} | n | \chi(t_1) \rho_{\beta}^{1/2}) - G_c(t_2, t_1) n_{\beta} \\
\label{eq:boundary-size-by-bulk-approximation}
&\approx \frac{\Delta b_{\Delta}}{4^{\Delta} \alpha_S}
\left( \frac{\beta \mathcal{J}}{2 \pi} \right)^{1 - 2 \Delta}
B_{\Delta} (\cosh \rho_0)^{2 \Delta} \langle \psi(x_2)_{j} (S + J_1 + J_2) \psi^{\dagger}(x_1)_{j} \rangle
\end{align}
where \(N_{\Delta, 1}\) is defined in Appendix \ref{sec:bulk-fermion-2pt}. In the Euclidean signature, the term in the derivative of the coordinates that gives rise to \(J_1\) is 
\begin{equation*}
\left. \frac{d \rho}{d\lambda} \right|_{\lambda = \pi} \partial_{\rho}(X_{>} - X_{<})
= - \frac{1}{\pi} (X_{>} - X_{<}).
\end{equation*}
Thus the contribution of this term to the size matrix element is actually just a constant, and we can replace \(J_1 \to - \frac{2 \Delta}{\pi}\). Its contribution to the boundary size is subleading in \(\epsilon\) (since it gives a contribution to the matrix element that is proportional to the two-point function). The \(J_2\) generator is not as simple, but its contribution also vanishes when \(t_2 = t_1\).

We then find that, to leading order in $\beta \mathcal{J} \gg 1$, the boundary size is proportional to the expectation of \(E - B\). In this sense we have given another derivation of the similar result in \cite{Lin:2019qwu}. One important difference is that we have also computed the ``off-diagonal matrix elements'' of \(n\), namely expectations like \((\chi(u_2) \rho_{\beta}^{1/2} | n | \chi(u_1) \rho_{\beta}^{1/2})\) where \(u_2 \ne u_1\). These will be essential for the computation of bulk operator size, as will be discussed in next section.

\section{Size of bulk fields}
\label{sec:size-of-bulk-fields}

In order to use boundary CFT computations to determine the size of ``bulk'' operators, we use an explicit construction of certain bulk operators as superpositions of boundary operators of various size. After describing our construction, we present some general properties of bulk size for SYK-type models.

\subsection{The explicit construction of bulk fields}
\label{sec:bulk-reconstruction-description}
The method for constructing bulk fields we pursue is analogous to that first worked out by Hamilton, Kabat, Lifschytz, and Lowe (HKLL) for certain quantizations of scalar~\autocite{Hamilton:2005ju, Hamilton:2006az, Hamilton:2006fh} and higher-spin fields~\autocite{Kabat:2012hp}. There are different ways to understand this procedure; here, we take an approach that makes explicit corrections due to interactions.

Consider a \(d\)-dimensional CFT with a bulk dual, with a spinor field \(\chi\) of dimension \(\Delta\). We work in the limit of large \(N\) and strong CFT coupling, so the dual \(\AdS_{d+1}\) theory is weakly coupled. Sources (of dimension \(d - \Delta\)) for the boundary field \(\chi\) correspond to boundary conditions for a bulk fermion \(\psi\). The fluctuating modes of \(\psi\) in the absence of sources, when taken near the boundary and appropriately scaled, behave as a fermion of dimension \(\Delta\) and are identified with \(\chi\). Explicitly, if \(z\) is some coordinate that approaches zero near the conformal boundary of \(\AdS\) (and \(x\) are the remaining coordinates),
\begin{equation}
\label{eq:bulk-boundary-limit-operator-eq}
\lim_{z \to 0} \psi(x, z) \leftrightarrow z^{\Delta} \chi(x).
\end{equation}
The behaviour of \(\chi\) can then distinguish different ways of approaching the boundary. Our main example will be a \(d=1\) model where the boundary lies at constant Rindler \(\rho\) coordinate in \(\AdS_2\). In this case, the explicit expression is
\begin{equation*}
\lim_{\rho \to \infty} \psi(t_R, \rho) \leftrightarrow (\sech \rho)^{\Delta} \chi(x).
\end{equation*} 

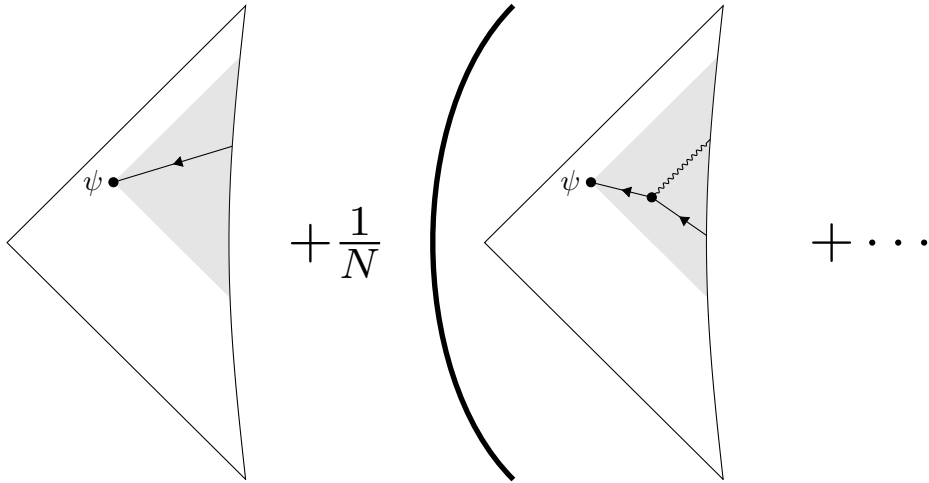
\begin{figure}[thb]
  \centering
  \begin{tikzpicture}[
    scale=2,
    fermi-prop/.style={decoration={markings, mark=at position 0.5 with {\arrow{Triangle}}},postaction=decorate},
    scalar-prop/.style={decorate,decoration={snake,amplitude=0.7,segment length=2.5,pre=lineto,pre length=2,post length=2,post=lineto}},
    field-point/.style={fill=black, radius=0.03cm}]

    \def\boundrho{2.9}
    \pgfmathsetmacro\coshboundrho{0.5*(e^\boundrho + e^(-\boundrho))}
    \def\rightrindlerboundpath{plot[smooth, domain=-pi/2:pi/2]
      ({pi / 180 * acos(cos(\x * 180 / pi) / \coshboundrho)}, {\x})}
    \foreach \patchlabel in {0, 1} {
      \tikzset{xshift={\patchlabel*2*pi/2*1cm}}

      \path [name path=right-rindler-boundary] \rightrindlerboundpath;
      \draw [name path=rindler-boundary] \rightrindlerboundpath
      -- (0, 0) -- (pi/2, -pi/2);

      \coordinate (bulk point) at (0.7, 0.4);

      \path [name path=lower-lc] let \p1 = (bulk point) in (\p1) -- ($(pi/2,\y1 + \x1 - pi/2*1cm)$);
      \path [name path=upper-lc] let \p1 = (bulk point) in (\p1) -- ($(pi/2, \y1 + pi/2*1cm - \x1)$);
      \path [name intersections={of=right-rindler-boundary and lower-lc,by=lower-lc-int}];
      \path [name intersections={of=right-rindler-boundary and upper-lc,by=upper-lc-int}];

      \begin{scope}
        \clip \rightrindlerboundpath -- (0, 0) -- (pi/2, -pi/2);
        \fill [gray,opacity=0.2] (bulk point) -- (upper-lc-int) -- (lower-lc-int) -- cycle;
      \end{scope}

      \node [left] at (bulk point) {$\psi$};
      \draw [field-point] (bulk point) circle;

      \ifthenelse{\patchlabel=0}{
        \begin{pgfinterruptboundingbox}
          \path [name path=bulk-propagator-outline] (bulk point) -- ++(17:pi/2);
          \draw [name intersections={of=right-rindler-boundary and bulk-propagator-outline}, fermi-prop]
          (intersection-1) -- (bulk point);
        \end{pgfinterruptboundingbox}
      }{
        \coordinate (interaction point) at (1.1, 0.3);
        \draw [field-point] (interaction point) circle;
        \draw [fermi-prop] (interaction point) -- (bulk point);
        \begin{pgfinterruptboundingbox}
          \path [name path=fermi-propagator-outline]
          (interaction point) -- ++(-35:pi/2);
          \draw [name intersections={of=right-rindler-boundary and
            fermi-propagator-outline}, fermi-prop]
          (intersection-1) -- (interaction point);

          \path [name path=scalar-prop-outline]
          (interaction point) -- ++(45:pi/2);
          \draw [name intersections={of=right-rindler-boundary and
            scalar-prop-outline}, scalar-prop]
          (intersection-1) -- (interaction point);
        \end{pgfinterruptboundingbox}
      }
    }
    \node [scale=2,right=4] (sum) at ($(pi/2, 0)$) {$ + \frac{1}{N}$};
    \draw[line width=2] (sum.east) ++(.7,-pi/2) .. controls +(135:1) and +(-135:1) .. ++(0, pi);
    \node [scale=2,right=4] at (5, 0) {$+ \cdots$};
  \end{tikzpicture}
  \caption{Illustration of the perturbative HKLL reconstruction, in the right Rindler wedge of $\ads_2$. We show a diagram that appears at lowest, and at next-to-lowest order in the interaction. We will focus on the lowest order contribution; our kernels only have support on the right boundary, so to this order the fermion is reconstructed only from boundary operators inside the right light cone (indicated in grey). In the picture with the spacelike propagator $G_F$ as in \eqref{eq:coordinate-invariant-reconstruction}, all the interaction vertices must be contained in the gray spacelike separated region. Fermion propagators are shown with a solid line, and the propagator of some putative scalar field interacting with the fermion is drawn with a wavy line.}
  \label{fig:perturbative-hkll}
\end{figure}

When the fermion is weakly interacting, we have the approximate equation
\begin{equation*}
(\slashed{\nabla} - m) \psi(x, z) \simeq 0.
\end{equation*}
This, in addition to the holographic principle, inspires us to look for a bispinor
\begin{equation*}
(\slashed{\nabla} - m) G_F(x, x') 
= G_F(x, x') (\overleftarrow{\slashed{\nabla}} - m)
= \frac{\delta^{d+1}(x -x')}{\sqrt{-g}}
\end{equation*}
with support only for spacelike separated \(x, x'\). Then we have, for \emph{any} spinor \(\psi(x)\) on a \(d+1\)-manifold \(M\) which we take to assume the boundary value \(\psi(x) \to z^{\Delta} \chi(x)\),
\begin{align}
\label{eq:coordinate-invariant-reconstruction}
\psi(x) &= \int_M d^{d+1} x' \sqrt{-g} (G_F(x, x') (\overleftarrow{\slashed{\nabla}} - m)) \psi(x') \\
&= \int_{\partial M} d(\partial M) (- \sqrt{-g} (z')^{\Delta} G_F(x, x') \slashed{N}) \chi(x') + \int_M d^{d+1} x' G_F(x, x') (\slashed{\nabla} - m) \psi(x')
\end{align}
where \(N\) is the outward-point normal vector to the boundary \(\partial M\), and \(\slashed{N} = N^{\mu} \Gamma_{\mu}\). This expansion provides a perturbative (in the interaction) diagrammatic approach to computing the bulk operator \(\psi(x)\) from boundary data; these will be bulk Witten diagrams with propagators replaced by ones like \(G_F\) with spacelike support. We illustrate this diagrammatic expansion in Figure~\ref{fig:perturbative-hkll}. Since we are in the large \(N\) limit, which suppresses interactions, we focus only on the first term in this expansion. We are assuming that in this limit, we can ignore the contribution of the interaction vertices to the bulk fermion. Regardless of the contribution of this term, as long as the boundary field \(\chi\) is nearly conformal, keeping just the first term gives an approximately local bulk field.

In special coordinate systems in the free limit, there is a more direct way to understand the HKLL procedure (this is the fermionic version of the ``mode sum'' approach taken in the original work). Essentially, the Fourier transform of the reconstruction kernel is the operator \(F_{\omega}\) that takes constant spinors to solutions of the Dirac equation and is an eigenfunction under the flow by the time coordinate \(t\), normalized so that as the coordinate \(z \to 0\), the dependence of \(F_{\omega}\) on \(z\) and \(t\) becomes \(F_{\omega} \to z^{\Delta} e^{- i \omega t}\). Then in Fourier space, the reconstruction happens simply by multiplying the boundary creation and annihilation operators at each momentum by the appropriate function of frequency. We give more details in Appendix \ref{sec:light-fermions}. In particular, we will use the fact that, for time-translation invariant quadratic expectations of fermions on the boundary, the Fourier transform of a bulk function in these special coordinates is a product of the Fourier transforms of the boundary function and the kernel.

In Appendix \ref{sec:light-fermions} and \ref{sec:reconstruction-kernel-details} we find concrete position-space expressions for these kernels for $\ads_2$. There are two cases to consider, depending on the sign of $\Delta - 1 / 2$, where $\Delta$ is the boundary spinor dimension. The simpler case is $\Delta > 1/2$, where we show that there exists a ``smearing kernel'' \(K_{\Delta}(x, z | y)\) such that, to leading order in \(N\) and at strong coupling (so that \(\chi(y)\) is nearly conformal and free),
\begin{equation}
\label{eq:large-delta-simple-reconstruction}
\psi_0(x, z) \equiv \int d^d y K_{\Delta}(x, z | y) \chi(y)
\end{equation}
behaves exactly as a free fermion on \(\AdS_{d+1}\) of mass \(|m| = \Delta - d/2\). This smearing function is indeed supported on points of the conformal boundary such that \((x, z)\) is space-like separated from \((y, z')\) as \(z' \to 0\). For example, in the \(\AdS_2\) Rindler coordinate (we do not normalize the kernels in any particular way, since our discussion will not depend on the normalization),
\begin{align}
\label{eq:large-delta-simple-rindler-kernel}
K_{\Delta}(t_R, \rho) &= 
\frac{1 + e^{\rho} \tanh (t_R/2) \gamma^2}{\sqrt{1-e^{2 \rho} \tanh^2 \left(\frac{t_R}{2} \right)}}
\left(
\cosh \rho - \sinh \rho \cosh t_R
\right)^{\Delta - 1} v_{-1} \Theta((t_R, \rho) \text{ spacelike to }0) \\
\gamma^1 v_{-1} &= - v_{-1} \\
K_{\Delta}(t_R, \rho | u) &= K_{\Delta}(t_R - u, \rho),
\end{align}
with gamma matrices \(\{\gamma^j, \gamma^k\} = \eta^{jk}\). Note that although there is only a single component fermion on a \(d=1\) boundary, the smearing kernel has two components for the two bulk fermions. Then, given a boundary size kernel \(\mathcal{G}^{\partial}_{\mu}(y, y')\), we can simply compute
\begin{align}
\label{eq:large-delta-simple-bulk-size-kernel}
\mathcal{G}^B_{\mu}(x, z, x', z')
&= \frac{( 0 |\mathcal{T}[e^{- \left(\int_0^{\beta} d\tau H^L(\tau) + \delta(\tau - \beta/2) \mu n(\tau)\right)} \psi^L_j(y) \psi^L_j(y')]| 0 )}
{( 0 |\mathcal{T}[e^{- \left(\int_0^{\beta} d\tau H^L(\tau) + \delta(\tau - \beta/2) \mu n(\tau)\right)}]| 0 )} \\
  \label{eq:explicit-bulk-size-from-kernels}
&= \int d^d y d^d y' K_{\Delta}(x, z|y)^{\dagger}_j K_{\Delta}(x', z'|y')_j \mathcal{G}^{\partial}_{\mu}(y, y').
\end{align}
The diagonal entries \(\mathcal{G}^B_{\mu}(x, z, x, z)_{jj}\) will be the generating function for the relative size distribution of \(\psi(x, z)\), as measured in terms of \(\chi(y)\) at \(t_0\).

In the SYK model, the boundary fermions have \(\Delta = 1/q < 1/2\), so the simpler kernel above does not apply. For a fermion of mass \(m\) in the bulk, the Dirac equation admits solutions that behave as \(z^{\Delta_{\pm}}\) for \(z \to 0\), where $\Delta_{\pm} = \frac{d}{2} \pm |m|$. In fact, for \(|m| < d/2\) there are two inequivalent ways to quantize a free fermion with no boundary sources, distinguished by their boundary behaviour. Since these quantizations have different behaviour (for example near the boundary), the smearing kernels must be different.  We give more details on derivations of this kernel in Appendices~\ref{sec:light-fermions} and \ref{sec:reconstruction-kernel-details}, and summarize the important points here. The simplest approach turns out to be to use the analytic form of the kernel \eqref{eq:large-delta-simple-rindler-kernel} (except changing \(v_{-1} \to v_1\) if we would like to keep the mass positive in the Dirac equation), but give a prescription for handling the non-integrable divergences in the kernel on the light cone for \(\Delta < 1/2\). One method is to use the analyticity of the free bulk modes in \(\Delta\), so that when integrating against analytic functions we just use a contour that analytically continues from the \(\Delta > 1/2\) case. Another possibility is to write the kernel as a linear differential operator that does not depend on time, acting on a function with integrable divergences. One way to accomplish this is
\begin{align*}
K_{\Delta}(t_R, \rho) &=
e^{- (\Delta + 1) \rho} \\
&\qquad\left[
(e^{\rho} + \gamma^2) \left( \slashed{\nabla} + (\Delta + \frac{1}{2}) (\csch \rho \gamma^0 - \coth \rho \gamma^1) - (\Delta + 1) \gamma^1 \right) \right. \\
&\qquad\qquad\left.
+ e^{\rho} \gamma^1
\right] \frac{e^{\rho}}{2} I(t_R, \rho) \Theta(\text{spacelike to } 0) v_1 \\
I(t_R, \rho) &= e^{- t_R / 2} \cosh^{2\Delta - 1} \left( \frac{t_R}{2} \right) (1 - x^2)^{\Delta - 1/2} \\
x &= e^{\rho} \tanh \left( \frac{t_R}{2} \right), v_1 = \gamma^1 v_1 \\
K_{\Delta}(t_R, \rho | u) &= K_{\Delta}(t_R - u, \rho);
\end{align*}
in this expression, the derivatives only act on \(t_R\), not \(u\). For \(u\) away from the light cone of \((t_R, \rho)\), we can evaluate the derivatives and find exactly \eqref{eq:large-delta-simple-rindler-kernel}, except with \(v_{-1}\) replaced by \(v_1\). The prescription is to regulate divergences by formally pulling the derivatives out of integrals against the kernel, and evaluating the derivatives on the now convergent integrals. More details on the different regularizations and quantizations can be found in Appendix \ref{sec:light-fermions}. The important point is that even in the case that regularization is required, the reconstruction is explicitly supported on points spacelike separated to the bulk point. We also point out that for the \(\Delta < 1/2\) case, the kernel diverges at the light cone, and there is a large weight for operators ``as late as possible'' in boundary time.

\subsection{Bulk size in the low temperature limit}
\label{sec:bulk-size-at-strong-coupling}

From the full boundary size matrix elements at low energy in \eqref{eq:boundary-size-bulk-generators}, and a perturbative definition of bulk operators, we proceed to compute the size of the bulk fermions. A schematic formula for the size of the bulk fermions constructed by HKLL in terms of the boundary Majorana fermions is given by
\begin{align}
n_{\beta}[\psi_j(x)]
&=
\frac{\int K_{\Delta}\left(x | u_2\right)_j K_{\Delta}\left(x | u_1\right)_j
\left[\left(\chi(u_2) \rho_{\beta}^{1/2} \right| n \left| \chi(u_1) \rho_{\beta}^{1/2}\right) - G_c(u_2, u_1) n_{\beta}\right]}
{\int K_{\Delta}(x | u_2)_j K_{\Delta}(x | u_1)_j G_c(u_2, u_1)} \\
\label{eq:bulk-size-formal-reconstruction}
&= 
\frac{\left(\psi(x)_j \rho_{\beta}^{1/2} \right| n \left| \psi(x)_j \rho_{\beta}^{1/2}\right) - \left\langle \psi(x)_j \psi(x)_j^{\dagger} \right\rangle n_{\beta}}{\left\langle \psi(x)_j \psi(x)_j^{\dagger} \right\rangle}
\end{align}
where \(\left\langle \psi(x)_j \psi(x)_j^{\dagger} \right\rangle\) is the covariant \(\ads_2\) fermion two-point function discussed in Appendix \ref{sec:bulk-fermion-2pt}. Note that as written, this formula does not seem to depend on the choice of coordinate, but does depend on the choice of vielbein used to define the kernel \(K_{\Delta}\) (since we are taking particular components).

The \(\ads_2\) fermion two-point function appears in the strong coupling limit since integration of the kernel \(K_{\Delta}\) against a conformal boundary two-point function gives the bulk function by construction. Likewise, ``expectations'' (i.e. expressions like \(\langle \psi(x) E \overline{\psi}(x') \rangle\)) of bulk symmetry generators are given by integrals against \(K_{\Delta}\) of expectations of boundary symmetry generators, since the reconstruction acts at the operator level.

To compute the numerator in \eqref{eq:bulk-size-formal-reconstruction}, we treat the three terms in \eqref{eq:size-matrix-element-generators} separately. The \(S = E - B\) generator has no boundary time dependence, so we can simply integrate over the two boundary fields separately before taking the expectation value, and find that this term becomes the bulk expectation of the same generator \(S\). As discussed at the end of Section \ref{sec:lowt-syk-boundary-size}, we can make the replacement \(J_1 \to - \frac{2 \Delta}{\pi}\), and so can make that replacement in the bulk as well.

At this point, we need to address the question of regulating \eqref{eq:bulk-size-formal-reconstruction}. In principle, a UV regulation in the boundary theory means that reconstructed bulk operators will have the singularities in their two-point functions smeared out as well, so we can take the two bulk points exactly equal in \eqref{eq:bulk-size-formal-reconstruction}. One way to understand the effect of a UV regulator is to regulate the boundary conformal two-point functions by an \(i \epsilon\) prescription (in the SYK model, we should take \(\epsilon \sim \pi / \beta \mathcal{J}\)). Since the kernel \(K_{\Delta}\) is time translation invariant, an \(i \epsilon\) regulation of a conformal boundary correlator is the same as splitting the bulk points by \(i \epsilon\) in the time coordinate, and keeping the boundary theory exactly conformal. Note that we have introduced some additional dependence on the coordinate choice. When we consider bulk points such that \(\cosh \rho \ll 1 / \epsilon\), the two points are split by a small (Euclidean) geodesic distance, so reconstructed bulk quantities at such coordinates will be dominated by the short distance divergence in the true bulk functions. Importantly, for coordinates such that \(\cosh \rho \gg 1 / \epsilon\) the Euclidean geodesic distance is large, and the bulk correlation functions become conformal. Thus in the large \(\rho\) limit, the bulk size is simply the boundary size, and approximated by an expectation of the symmetry generator \(S=E-B\).

It remains to understand the contribution of the \(J_2\) term in the bulk. Since it is time translation invariant, we can directly compute the Fourier transform of its bulk contribution. We will need the Fourier transform
\begin{equation*}
\int du e^{i \omega u} \left( \sin \left(\frac{i u + \epsilon}{2}\right) \right)^{- 2 p}
= 4^{p} e^{(\pi - \epsilon) \omega} B(p + i \omega, p - i \omega).
\end{equation*}
Since the bulk size matrix element is given by dividing bulk expectations of generators by the bulk two-point function, only the high-frequency behaviour contributes as we take the separation between bulk points to zero. The contribution of the \(J_2\) term becomes \(\frac{1}{\pi}(1 + \omega \partial_{\omega}) G(\omega)\), where \(G(\omega)\) is the Fourier transform of the boundary two-point function.\(G\) decays exponentially for large negative \(\omega\), but for large positive \(\omega\) it has power law behaviour \(\sim \omega^{2 \Delta - 1}\). Consequently, at large frequency this term becomes \(\sim \frac{2 \Delta}{\pi} G(\omega)\), and therefore contributes a constant \(\frac{2 \Delta}{\pi}\) to the bulk size. Therefore we find that the bulk size (such that \(\cosh \rho \ll 1 / \epsilon\)) is given by the bulk quantity
\begin{equation}
\label{eq:psi-size-bulk-prop-argument}
n_{\beta}[\psi(t_R, \rho)_j] 
\propto \frac{\langle \psi(t_R - i \epsilon, \rho)_j (E - B) \psi(t_R, \rho)^{\dagger}_j \rangle}{\langle \psi(t_R - i \epsilon, \rho)_j \psi(t_R, \rho)^{\dagger}_j \rangle}.
\end{equation}

Using \eqref{eq:boundary-size-by-bulk-approximation} to fix the normalization constants (i.e. units of size), we conclude that the bulk size both for small and large \(\rho\) is well-approximated by
\begin{equation}
  \label{eq:bulk-and-boundary-size-approx-normed}
  n[\psi(x)_j]_{\beta} \approx \frac{\Delta b_{\Delta}}{4^{\Delta} \alpha_S} \left( \frac{\beta \mathcal{J}}{2 \pi} \right)^{1-2\Delta}
  \frac{\langle \psi(t_R - i \epsilon, \rho)_j (E - B) \psi(t_R, \rho)^{\dagger}_j \rangle}
  {\langle \psi(t_R - i \epsilon, \rho)_j \psi(t_R, \rho)^{\dagger}_j\rangle},
\end{equation}
where \(\epsilon \sim \pi / \mathcal{\beta \mathcal{J}}\) determines what we mean by ``small'' and ``large'' \(\rho\), and \(\nu\) is some numerical constant that can be used to fix units of size. Calling the vector field \(V\) that generates the symmetry of \(\ads_2\) associated to \(E - B\), we have \(\langle \psi(x)_j (E - B) \psi(x')^{\dagger}_j \rangle = i V^{\mu} \nabla_{\mu} G_{\psi}(x, x') \gamma^{0}\), where \(G_{\psi}(x, x') = \langle \psi(x) \overline{\psi(x')} \rangle\) is a covariant two-point function. As mentioned, the behaviour at large \(\rho\) is just the boundary size \eqref{eq:boundary-size-behaviour}. We compute the full expression for this expectation of generators in Appendix \ref{sec:bulk-fermion-2pt}, but here we note the simple behaviour in the limit \(\epsilon \to 0\),
\begin{align}
  \label{eq:deep-bulk-size-approximation}
\frac{\langle \psi(t_R - i \epsilon, \rho)_j (E - B) \psi(t_R, \rho)^{\dagger}_j \rangle}
{\langle \psi(t_R - i \epsilon, \rho)_j \psi(t_R, \rho)^{\dagger}_j\rangle}
  &\xrightarrow[\epsilon \to 0]{} \frac{\coth \rho \cosh t_R - 1}{\epsilon} \\
  &= \epsilon^{-1} \left( \frac{\sin \sigma \cos t_G}{\sin^2 \sigma - \sin^2 t_G} - 1 \right)
\end{align}
where we have also given the limit in global coordinates $\sigma, t_G$. At fixed $\epsilon$, this is an accurate approximation to the size for $\cosh \rho \ll 1 / \epsilon$. The behaviour of this function in the Rindler wedge is shown in Figure~\ref{fig:deep-bulk-size}.
\pgfplotsset{
  rindler global/.style={
    xlabel=$\sigma$,
    ylabel=$t_G$,
    xtick={0},
    xticklabels={0},
    ytick={pi/2},
    yticklabels={$\pi/2$},
    xmin=0,
    xmax=pi/2,
    ymin=0,
    ymax=pi/2,
    colormap name=viridis,
    y label style={at={(ticklabel* cs:0.5)}},
    x label style={at={(ticklabel* cs:0.5)}},
  }
}
\makeatletter \def\pgfplotscolormappdfmax{1} \def\pgfplotscolormappdfmax@inv{1000}
\makeatother
\begin{figure}[thbp]
  \centering
  \begin{tikzpicture}
    \begin{axis}[
      declare function={
        epsl=0.01;
        epsr=0.05;
        zminval=-7.5;
        tG(\x,\y)=(\y)*(pi/2 - (\x));
        sigma(\x,\y)=(\x) + (\y)*(pi/2 - (\x));
        size(\x,\y)=ln((sin(deg(sigma(\x, \y))) * cos(deg(tG(\x,\y))))/((sin(deg(sigma(\x,\y))))^2-(sin(deg(tG(\x,\y))))^2)-1);
      },
      rindler global,
      contour/labels=false,
      contour/number=21,
      ztick=\empty,
      zmin=zminval,
      zlabel={$\log$ bulk size},
      domain={epsl:(pi/2-epsr)}, domain y={0:1}
      ]
      % levels={-5,-4,-3,-2,-1,0,1,3}
      \addplot3
      [contour gnuplot,
      contour/draw color=black,
      z filter/.expression={zminval},
      update limits=false]
      ({sigma(\x,\y)}, {tG(\x,\y)},
      {size(\x,\y)});
      \addplot3
      [surf, samples=30, shader=interp]
      ({sigma(\x,\y)}, {tG(\x,\y)},
      {size(\x,\y)});
      \addplot3
      [contour gnuplot,
      contour/draw color=white]
      ({sigma(\x,\y)}, {tG(\x,\y)},
      {size(\x,\y)});
    \end{axis}
  \end{tikzpicture}
  \caption{Bulk size in the limit $\cosh \rho \ll 1 / \epsilon$ (in arbitrary size units; note that in this limit, the bulk fermion size of both components has the same behaviour). The plot is shown in global coordinates $\sigma, t_G$ over the part of the Rindler wedge $t_R > 0$. As discussed in Section~\ref{sec:numerics-at-large-q}, this remains a good approximation to the bulk size in the presence of finite $\beta \mathcal{J}$ corrections on the boundary.}
  \label{fig:deep-bulk-size}
\end{figure}
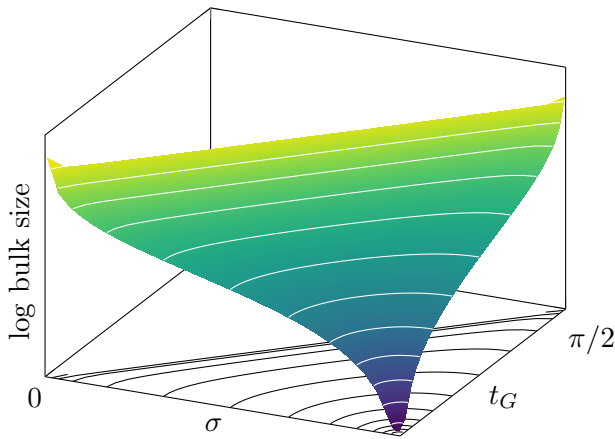

\subsection{Numerics at large $q$}
\label{sec:numerics-at-large-q}

The boundary size distribution is also known at large $q$, for all values of the coupling~\autocite{qi2019quantum}:
\begin{multline}
  \label{eq:large-q-boundary-size-distribution}
  \mathcal{G}_{\mu}\left(\frac{\beta^+}{4} + i u, \frac{\beta^-}{4} + iu'\right) = \\
  \frac{e^{- \mu} G_{\mu}(u - u')}{
    \left( 1 + \frac{1 - e^{- q \mu}}{2} \left( \frac{\mathcal{J}}{\alpha_{\mu}} \right)^2 G_{\mu}(u - u')^{q/2} (\cosh (\alpha_{\mu}(u + u')) - \cosh(\alpha_{\mu} (u - u') - i\epsilon)) \right)^{2/q}}
\end{multline}
where
\begin{equation}
  G_{\mu}(u) = \left( \frac{\sin \gamma_{\mu}}{\sin (\gamma_{\mu} + i \alpha_{\mu} u)} \right)^{2/q},
\end{equation}
and \(\alpha_{\mu}\) and \(\gamma_{\mu}\) satisfy
\begin{equation}
  \sin \gamma_{\mu} = \frac{\alpha_{\mu}}{\mathcal{J}} \text{ and } \sin \left( \frac{\alpha_{\mu} \beta}{2} + 2 \gamma_{\mu} \right) = e^{- q \mu} \sin \left( \frac{\alpha_{\mu} \beta}{2} \right).
\end{equation}
At low temperature $N \gg \beta \mathcal{J} \gg 1$, our discussion above applies. To help further understand the effect of finite $\beta \mathcal{J}$ corrections on the boundary, we numerically compute the bulk size using \eqref{eq:explicit-bulk-size-from-kernels} directly\footnote{If one interprets finite $\beta \mathcal{J}$ corrections as affecting the location of the boundary, it is natural to wonder if we should correct the reconstruction kernel as well. For fermions, we cannot naively put the boundary at a finite location in \eqref{eq:coordinate-invariant-reconstruction} to include such corrections for the technical reason that it is only in the limit of a conformal boundary that the kernel is proportional to a projector, while we always only have a single fermion component on the boundary.}. We have to regulate divergent integrals against the kernel, and have checked that both methods described in Appendix~\ref{sec:poincare-coordinate-modes} agree; details on practically useful numerical versions of these schemes are given in Appendix~\ref{app:numerics}.

\begin{figure}[thbp]
  \centering
  \def\plotwidth{2.4in}
  \pgfplotsset{
    colormap/blackwhite,
    projection/.style={
      opacity=0.3,
      scatter=false,
      color=black,
    },
    every axis plot/.style={
      only marks,
      scatter,
      z buffer=sort,
      mark size={.7},
    },
    every axis/.append style={
      rindler global,
      width=\plotwidth,
      height=\plotwidth,
    }
  }
  \begin{subfigure}[b]{0.45\linewidth}
    \begin{tikzpicture}
      \begin{axis}[zlabel={$\log \frac{\text{Numerical size}}{\text{Approximation \eqref{eq:deep-bulk-size-approximation}}}$}]
        \addplot3[y filter/.expression={pi/2}, projection] table [x=sigma,y=tG,z=boundary_comp] {plot_data/high_t_comparison.dat};
        \addplot3[x filter/.expression={0}, projection] table [x=sigma,y=tG,z=boundary_comp] {plot_data/high_t_comparison.dat};
        \addplot3 table [x=sigma,y=tG,z=boundary_comp] {plot_data/high_t_comparison.dat};
      \end{axis}
    \end{tikzpicture}
    \caption{Fermion component decaying slower near the boundary.}
    \label{fig:large-q-deep-bulk-approx-bndry}
  \end{subfigure}
  \begin{subfigure}[b]{0.45\linewidth}
    \begin{tikzpicture}
      \begin{axis}
        \addplot3[y filter/.expression={pi/2}, projection] table [x=sigma,y=tG,z=nonboundary_comp] {plot_data/high_t_comparison.dat};
        \addplot3[x filter/.expression={0}, projection] table [x=sigma,y=tG,z=nonboundary_comp] {plot_data/high_t_comparison.dat};
        \addplot3 table [x=sigma,y=tG,z=nonboundary_comp] {plot_data/high_t_comparison.dat};
      \end{axis}
    \end{tikzpicture}
    \caption{Fermion component decaying faster near the boundary.}
    \label{fig:large-q-deep-bulk-approx-bndry}
  \end{subfigure}
  \caption{Comparison of the approximation~\eqref{eq:deep-bulk-size-approximation} and numerical results for $q=1000$ and $\beta \mathcal{J} \approx 61$ ($\pi - \beta \alpha_{\mu = 0} = 0.1$). In particular, we show the logarithm of the ratio between these expressions for the two bulk fermion components. We show it both as a function of $t_G$ and $\sigma$, and as the profile seen from $\sigma = 0$ or $t_G = \pi/2$. The ratio is constant for a large portion of the bulk. The deviations near $\sigma = \pi/2$, $t_G = 0$ are significant for both components. There is an abrupt drop, not present in the simple approximation, in the size of the component decaying slower near the boundary as the boundary is approached that only appears as a line of points in this figure. We note that for the purpose of these plots, we have fixed a choice of size unit (an overall constant multiplying the numerical size).}
  \label{fig:large-q-deep-bulk-approx-comparison}
\end{figure}
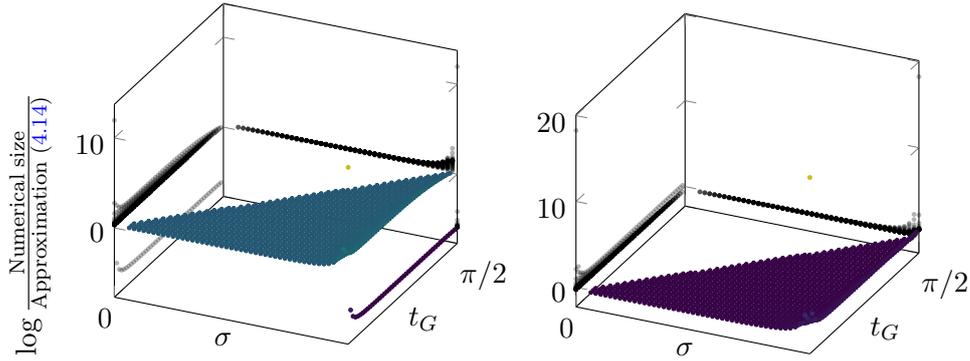

First, we note that even at relatively small coupling the approximation \eqref{eq:deep-bulk-size-approximation} captures both the qualitative and quantitative behaviour of size away from the boundary. We illustrate this by showing the logarithm of the ratio between the approximation and the numerical result for a relatively small coupling $\beta \mathcal{J} \approx 61$ in Figure~\ref{fig:large-q-deep-bulk-approx-comparison}. For stronger couplings, the agreement holds nearer to the boundary, as expected. In light of this, we will concentrate on the behaviour near the boundary, \(\rho \to \infty\), for the remainder of this section.

Here, the size of the component of the bulk field decaying faster near the boundary, in other words the field ``not present'' at the boundary, asymptotes to a constant size greater than \(n_{\beta}[\chi(t_R)]\), while the other component briefly levels off at this larger size, then rapidly drops to \(n_{\beta}[\chi(t_R)]\) as we go further towards the boundary. We refer to the two components, respectively, as the ``non-boundary'' and ``boundary'' components. Figure~\ref{fig:3d-rindler-size-plot} shows this behaviour for a particular temperature. The location of this rapid drop in the size is a function of \(\beta \mathcal{J}\), with lower temperatures pushing the location of the drop in size to larger \(\rho\). Some example sizes demonstrating this pattern are shown in Figures \ref{fig:fixed-tR0-size} and \ref{fig:fixed-tR1-size}. This suggests to identify the approximate location of the boundary with this drop. Further numerical evidence for this identification is that, once we find some \(\rho\) at some fixed \(t_R\) at which the bulk size of the ``boundary'' component approaches \(\tilde{n}_{\beta}[\chi(t_R)]\), the boundary value is approached at the same \(\rho\) for different times \(t_R\).

\tikzset{
  inflection-marker/.style={
    shape=star, fill=red, minimum size=5pt, inner sep=0, star point ratio=2.3
  }
}%
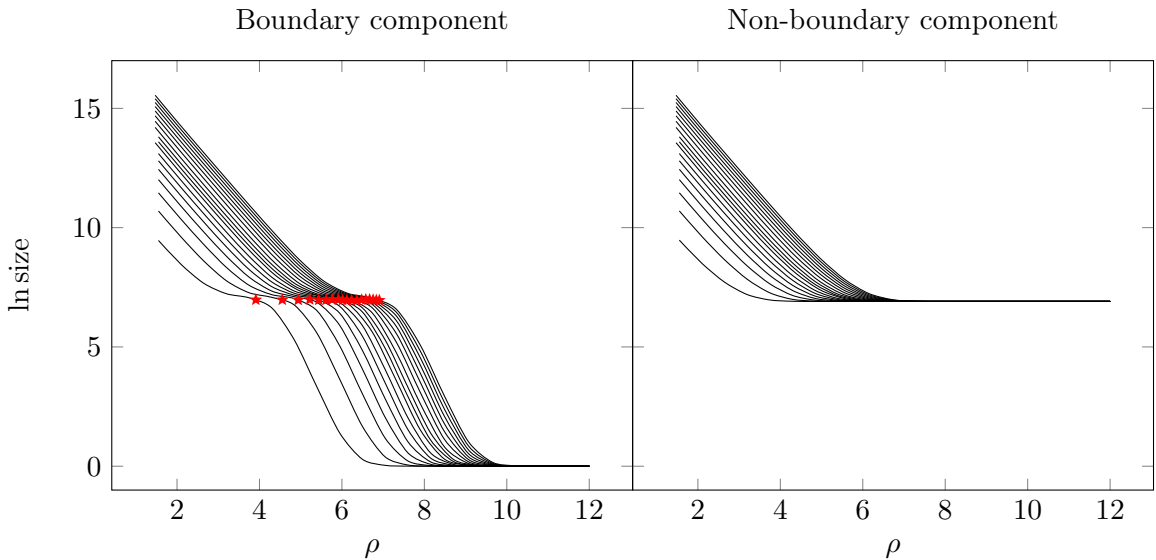
\begin{figure}[thbp]
\centering
\def\plotbetas{49.999999999999815,95.0000000000026,139.99999999998943,184.9999999999795,229.9999999999537,275.00000000003536,319.99999999985937,370.000000000044,410.0000000000568,454.99999999990945,510.0000000005339,579.9999999998295,649.9999999997694,719.9999999996419,789.999999998293,859.9999999996525,930.0000000013973,1000.0000000019079}
\pgfplotsset{
  discard unless/.style 2 args={
    x filter/.expression={
      x <= 1 ? nan : (\thisrow{#1} == #2 ? x : nan)
    },
    filter discard warning=false
  },
  fixtR/.style={
    ymin=-1,
    ymax=17
  },
}
\begin{tikzpicture}
  \begin{groupplot}[group style={group size=2 by 1, horizontal sep=0}, fixtR, xlabel={$\rho$}]
    \nextgroupplot[ylabel={$\ln \text{size}$}, title={Boundary component}]
    \foreach \curbeta in \plotbetas {
      \edef\temp{
        \noexpand\addplot[discard unless={beta}{\curbeta}, smooth, name path={size-path-\curbeta}] table [x={rho},y expr={ln(\noexpand\thisrow{size_bndry})}] {plot_data/rindler_tR0_size.dat};}
      \temp
    }
    \foreach \curbeta in \plotbetas {
      \edef\temp{
        \noexpand\path[name path={bound-vert}] ($(axis cs:{ln(\curbeta)},0)$ ) -- ++(axis direction cs:0, 20);
        \noexpand\node[inflection-marker, name intersections={of={size-path-\curbeta} and bound-vert}] at (intersection-1) {};}
      \temp
    }
    \nextgroupplot[yticklabels={}, title={Non-boundary component}]
    \foreach \curbeta in \plotbetas {
      \addplot[discard unless={beta}{\curbeta}, smooth] table [x={rho},y expr={ln(\thisrow{size_nonbndry})}] {plot_data/rindler_tR0_size.dat};
    }
  \end{groupplot}
  \end{tikzpicture}
\caption{\label{fig:fixed-tR0-size}Size at fixed \(t_R = 0\) for various \(\beta \mathcal{J}\) and \(q=1000\). Lines corresponding to higher $\beta \mathcal{J}$ lie to the right. We have indicated the intersection of the vertical line $\ln \beta \mathcal{J}$ with the respective size curve by a star.}
\end{figure}

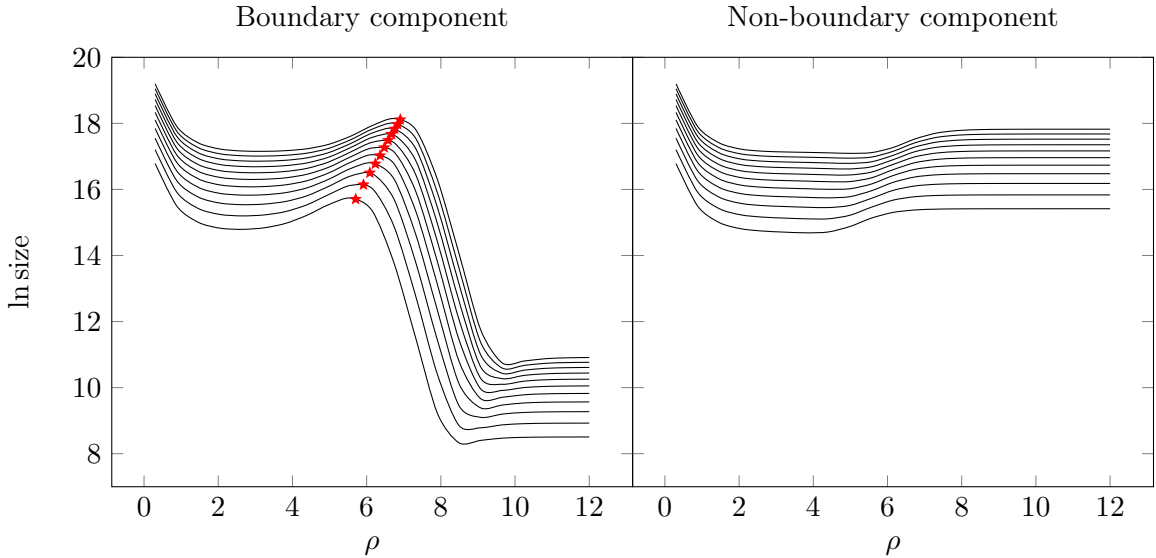
\begin{figure}[thbp]
  \centering
  \def\plotbetas{299.9999999998786,370.000000000044,439.99999999993753,510.0000000005339,579.9999999998295,649.9999999997694,719.9999999996419,789.999999998293,859.9999999996525,930.0000000013973,1000.0000000019079}
  \pgfplotsset{
    discard unless/.style 2 args={
      x filter/.expression={
        \thisrow{#1} == #2 ? x : nan
      }
    },
    fixtR/.style={
      ymin=7,
      ymax=20
    }
  }
  \begin{tikzpicture}
    \begin{groupplot}[group style={group size=2 by 1, horizontal sep=0}, fixtR, xlabel={$\rho$}]
      \nextgroupplot[ylabel={$\ln \text{size}$}, title={Boundary component}]
      \foreach \curbeta in \plotbetas {
        \edef\temp{
          \noexpand\addplot[discard unless={beta}{\curbeta}, smooth, name path={size-path-\curbeta}] table [x={rho},y expr={ln(\noexpand\thisrow{size_bndry})}] {plot_data/rindler_tR1_size.dat};}
        \temp
      }
      \foreach \curbeta in \plotbetas {
        \edef\temp{
          \noexpand\path[name path={bound-vert}] ($(axis cs:{ln(\curbeta)},0)$ ) -- ++(axis direction cs:0, 20);
          \noexpand\node[inflection-marker, name intersections={of={size-path-\curbeta} and bound-vert}] at (intersection-1) {};}
        \temp
      }
      \nextgroupplot[yticklabels={}, title={Non-boundary component}]
      \foreach \curbeta in \plotbetas {
        \addplot[discard unless={beta}{\curbeta}, smooth] table [x={rho},y expr={ln(\thisrow{size_nonbndry})}] {plot_data/rindler_tR1_size.dat};
      }
    \end{groupplot}

  \end{tikzpicture}
  \caption{\label{fig:fixed-tR1-size}Size at fixed \(t_R = 1\) for various \(\beta \mathcal{J}\) and \(q = 1000\). As in Figure~\ref{fig:fixed-tR0-size}, curves at lower temperature lie to the right, we have marked the intersection of the vertical line $\ln \beta \mathcal{J}$ with the respective size curve by a star.}
\end{figure}

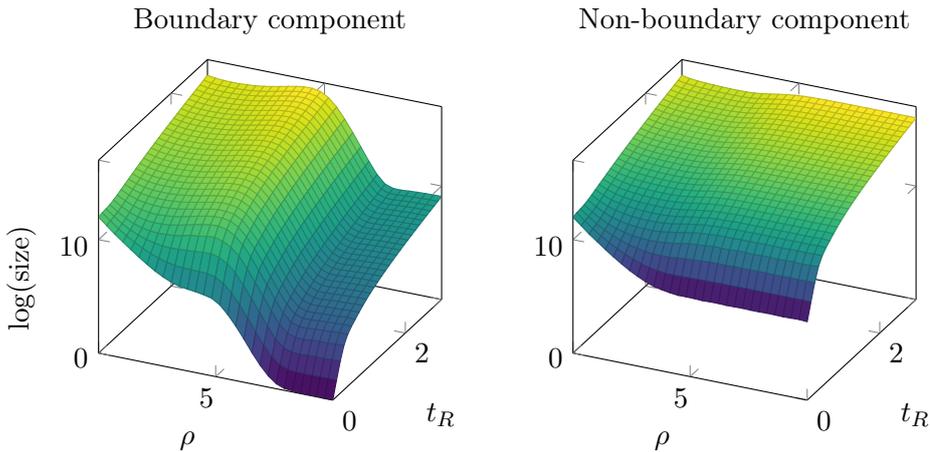
\begin{figure}[thbp]
\centering
\def\plotwidth{2.4in}
\pgfplotsset{
  every axis/.append style={
    width=\plotwidth,
    height=\plotwidth,
    xlabel=$\rho$,
    ylabel=$t_R$,
    colormap name=viridis,
    zmin=0, zmax=17}}
\begin{subfigure}[b]{0.45\linewidth}
  \begin{tikzpicture}
    \begin{axis}[zlabel={$\log (\text{size})$}, title={Boundary component}]
      \addplot3[surf]
      gnuplot [raw gnuplot] { set dgrid3d 30,30 splines; splot "plot_data/rindler_rhot_size_dep.dat" using 2:1:3;};
    \end{axis}
  \end{tikzpicture}
\end{subfigure}
\begin{subfigure}[b]{0.45\linewidth}
  \begin{tikzpicture}
    \begin{axis}[title={Non-boundary component}]
      \addplot3[surf]
      gnuplot [raw gnuplot] { set dgrid3d 30,30 splines; splot "plot_data/rindler_rhot_size_dep.dat" using 2:1:4;};
    \end{axis}
  \end{tikzpicture}
\end{subfigure}
\caption{\label{fig:3d-rindler-size-plot}Size for large \(\rho\) for \(q=1000\), \(\beta \mathcal{J} = 100\). The fast ``dip'' and subsequent saturation in size of the \(\gamma^1=+1\) component occurs at a fixed \(\rho\) for all \(t_R\), well after the \(\gamma^1=-1\) component has saturated. Note there is a finite range of \(\rho\) where it would appear both components are saturating to the same size.}
\end{figure}

\section{Conclusion}
\label{sec:conclusion}

In conclusion, we have studied the operator size distribution of bulk dual fermion of SYK model, using a combination of the HKLL formalism and SYK calculations. Our results provide an explicit proof of the relation between operator size and $\ads_2$ quantum number in the bulk. Operator size grows exponentially for operators deeper in the bulk, which therefore can be used as a measure of the bulk emergent spatial dimension. In higher dimensions, it is easier to see how operators deeper inside the bulk are more complicated, since they can only be reconstructed on a bigger region on the boundary\cite{almheiri2015bulk}. For a $0+1$-dimensional bulk theory, since there is no spatial locality on the boundary, it is more difficult to quantify the relation of emergent bulk spatial dimension with complexity and quantum error correction. The operator size distribution provides a useful tool to make progress along this direction.

There are many open questions along this direction. One question is whether there is an analog of the quantum error correction understanding of bulk locality in higher dimensions. How local is the bulk dual theory of SYK model in sub-AdS scale? How is sub-AdS locality related to the operator size distribution? It is also interesting to ask how to generalize the operator size measure and its dual interpretation to other models, such as the eternal traversable wormhole ({\it i.e.} global AdS$_2$) geometry that is dual to a pair of coupled SYK sites\cite{maldacena2018eternal}. Intuitively, when a fermion moves from one boundary to the other in the global AdS$_2$ geometry, one expects the operator size to increase and then decrease. The temperature dependent operator size measure (\ref{eq:effectivesize}) does not directly apply, because the two sites together could be at zero temperature. This suggests that a more general relation between operator size and bulk spatial dimension requires a modified operator size measure. 

\noindent{\bf Acknowledgement.} We would like to thank Yingfei Gu, Henry Lin, Douglas Stanford, Zhuo-Yu Xian and Ying Zhao for helpful discussion, as well as the anonymous JHEP reviewer for insightful comments. PZ would like to thank Yingfei Gu for bring his attention to the operator size of SYK models in the low temperature limit. This work is supported by the National Science Foundation Grant No. 1720504 (YL and XLQ), the Hertz Foundation (YL) and the Walter Burke Institute for Theoretical Physics at Caltech (PZ). This work is also supported in part by the
DOE Office of Science, Office of High Energy Physics, the grant DE-SC0019380 (XLQ).

\bibliography{syk_bulk_size_paper}

\providecommand{\href}[2]{#2}\begingroup\raggedright\begin{thebibliography}{10}

\bibitem{maldacena1999large}
J.~M. Maldacena, {\it {The Large N limit of superconformal field theories and
  supergravity}},  {\em Int. J. Theor. Phys.} {\bf 38} (1999) 1113--1133,
  [\href{http://xxx.lanl.gov/abs/hep-th/9711200}{{\tt hep-th/9711200}}].

\bibitem{sekino2008fast}
Y.~Sekino and L.~Susskind, {\it {Fast Scramblers}},  {\em JHEP} {\bf 10} (2008)
  065, [\href{http://xxx.lanl.gov/abs/0808.2096}{{\tt arXiv:0808.2096}}].

\bibitem{shenker2014black}
S.~H. Shenker and D.~Stanford, {\it {Black holes and the butterfly effect}},
  {\em JHEP} {\bf 03} (2014) 067,
  [\href{http://xxx.lanl.gov/abs/1306.0622}{{\tt arXiv:1306.0622}}].

\bibitem{larkin1969quasiclassical}
A.~Larkin and Y.~N. Ovchinnikov, {\it Quasiclassical method in the theory of
  superconductivity},  {\em Sov Phys JETP} {\bf 28} (1969), no.~6 1200--1205.

\bibitem{shenker2014multiple}
S.~H. Shenker and D.~Stanford, {\it {Multiple Shocks}},  {\em JHEP} {\bf 12}
  (2014) 046, [\href{http://xxx.lanl.gov/abs/1312.3296}{{\tt
  arXiv:1312.3296}}].

\bibitem{kitaev2014talk}
A.~Kitaev, {\it "hidden correlations in the hawking radiation and thermal
  noise", talk at breakthrough prize symposium,
  \url{https://www.youtube.com/watch?v=OQ9qN8j7EZI}}, .

\bibitem{hosur2016characterizing}
P.~Hosur and X.-L. Qi, {\it Characterizing eigenstate thermalization via
  measures in the fock space of operators},  {\em Physical Review E} {\bf 93}
  (Apr, 2016) [\href{http://xxx.lanl.gov/abs/1507.0400}{{\tt
  arXiv:1507.0400}}].

\bibitem{roberts2018operator}
D.~A. Roberts, D.~Stanford, and A.~Streicher, {\it {Operator growth in the SYK
  model}},  {\em JHEP} {\bf 06} (2018) 122,
  [\href{http://xxx.lanl.gov/abs/1802.0263}{{\tt arXiv:1802.0263}}].

\bibitem{qi2019quantum}
X.-L. Qi and A.~Streicher, {\it {Quantum Epidemiology: Operator Growth, Thermal
  Effects, and SYK}},  {\em JHEP} {\bf 08} (2019) 012,
  [\href{http://xxx.lanl.gov/abs/1810.1195}{{\tt arXiv:1810.1195}}].

\bibitem{qi2019measuring}
X.-L. Qi, E.~J. Davis, A.~Periwal, and M.~Schleier-Smith, {\it {Measuring
  operator size growth in quantum quench experiments}},
  \href{http://xxx.lanl.gov/abs/1906.0052}{{\tt arXiv:1906.0052}}.

\bibitem{xu2019locality}
S.~Xu and B.~Swingle, {\it {Locality, Quantum Fluctuations, and Scrambling}},
  {\em Phys. Rev. X} {\bf 9} (2019), no.~3 031048,
  [\href{http://xxx.lanl.gov/abs/1805.0537}{{\tt arXiv:1805.0537}}].

\bibitem{vermersch2019probing}
B.~Vermersch, A.~Elben, L.~M. Sieberer, N.~Y. Yao, and P.~Zoller, {\it {Probing
  scrambling using statistical correlations between randomized measurements}},
  {\em Phys. Rev. X} {\bf 9} (2019), no.~2 021061,
  [\href{http://xxx.lanl.gov/abs/1807.0908}{{\tt arXiv:1807.0908}}].

\bibitem{lucas2019operator}
A.~Lucas, {\it {Operator size at finite temperature and Planckian bounds on
  quantum dynamics}},  {\em Phys. Rev. Lett.} {\bf 122} (2019), no.~21 216601,
  [\href{http://xxx.lanl.gov/abs/1809.0776}{{\tt arXiv:1809.0776}}].

\bibitem{mousatov2019operator}
A.~Mousatov, {\it {Operator Size for Holographic Field Theories}},
  \href{http://xxx.lanl.gov/abs/1911.0508}{{\tt arXiv:1911.0508}}.

\bibitem{brown2018falling}
A.~R. Brown, H.~Gharibyan, A.~Streicher, L.~Susskind, L.~Thorlacius, and
  Y.~Zhao, {\it {Falling Toward Charged Black Holes}},  {\em Phys. Rev. D} {\bf
  98} (2018), no.~12 126016, [\href{http://xxx.lanl.gov/abs/1804.0415}{{\tt
  arXiv:1804.0415}}].

\bibitem{Susskind:2018tei}
L.~Susskind, {\it {Why do Things Fall?}},
  \href{http://xxx.lanl.gov/abs/1802.0119}{{\tt arXiv:1802.0119}}.

\bibitem{Susskind:2019ddc}
L.~Susskind, {\it {Complexity and Newton's Laws}},
  \href{http://xxx.lanl.gov/abs/1904.1281}{{\tt arXiv:1904.1281}}.

\bibitem{Lin:2019qwu}
H.~W. Lin, J.~Maldacena, and Y.~Zhao, {\it {Symmetries Near the Horizon}},
  {\em JHEP} {\bf 08} (2019) 049,
  [\href{http://xxx.lanl.gov/abs/1904.1282}{{\tt arXiv:1904.1282}}].

\bibitem{jackiw1985lower}
R.~Jackiw, {\it Lower dimensional gravity},  {\em Nuclear Physics B} {\bf 252}
  (1985) 343--356.

\bibitem{teitelboim1983gravitation}
C.~Teitelboim, {\it Gravitation and hamiltonian structure in two spacetime
  dimensions},  {\em Physics Letters B} {\bf 126} (1983), no.~1-2 41--45.

\bibitem{Gu:2017njx}
Y.~Gu, A.~Lucas, and X.-L. Qi, {\it {Spread of entanglement in a
  Sachdev-Ye-Kitaev chain}},  {\em JHEP} {\bf 09} (2017) 120,
  [\href{http://xxx.lanl.gov/abs/1708.0087}{{\tt arXiv:1708.0087}}].

\bibitem{Hamilton:2005ju}
A.~Hamilton, D.~N. Kabat, G.~Lifschytz, and D.~A. Lowe, {\it {Local bulk
  operators in AdS/CFT: A Boundary view of horizons and locality}},  {\em Phys.
  Rev.} {\bf D73} (2006) 086003,
  [\href{http://xxx.lanl.gov/abs/hep-th/0506118}{{\tt hep-th/0506118}}].

\bibitem{Hamilton:2006az}
A.~Hamilton, D.~N. Kabat, G.~Lifschytz, and D.~A. Lowe, {\it {Holographic
  representation of local bulk operators}},  {\em Phys. Rev.} {\bf D74} (2006)
  066009, [\href{http://xxx.lanl.gov/abs/hep-th/0606141}{{\tt
  hep-th/0606141}}].

\bibitem{Hamilton:2006fh}
A.~Hamilton, D.~N. Kabat, G.~Lifschytz, and D.~A. Lowe, {\it {Local bulk
  operators in AdS/CFT: A Holographic description of the black hole interior}},
   {\em Phys. Rev.} {\bf D75} (2007) 106001,
  [\href{http://xxx.lanl.gov/abs/hep-th/0612053}{{\tt hep-th/0612053}}].
  [Erratum: Phys. Rev.D75,129902(2007)].

\bibitem{Kabat:2012hp}
D.~Kabat, G.~Lifschytz, S.~Roy, and D.~Sarkar, {\it {Holographic representation
  of bulk fields with spin in AdS/CFT}},  {\em Phys. Rev.} {\bf D86} (2012)
  026004, [\href{http://xxx.lanl.gov/abs/1204.0126}{{\tt arXiv:1204.0126}}].

\bibitem{Gross:2017hcz}
D.~J. Gross and V.~Rosenhaus, {\it {The Bulk Dual of SYK: Cubic Couplings}},
  {\em JHEP} {\bf 05} (2017) 092,
  [\href{http://xxx.lanl.gov/abs/1702.0801}{{\tt arXiv:1702.0801}}].

\bibitem{Kitaev:2017awl}
A.~Kitaev and S.~J. Suh, {\it {The soft mode in the Sachdev-Ye-Kitaev model and
  its gravity dual}},  {\em JHEP} {\bf 05} (2018) 183,
  [\href{http://xxx.lanl.gov/abs/1711.0846}{{\tt arXiv:1711.0846}}].

\bibitem{Maldacena:2016hyu}
J.~Maldacena and D.~Stanford, {\it {Remarks on the Sachdev-Ye-Kitaev model}},
  {\em Phys. Rev.} {\bf D94} (2016), no.~10 106002,
  [\href{http://xxx.lanl.gov/abs/1604.0781}{{\tt arXiv:1604.0781}}].

\bibitem{Bagrets:2016cdf}
D.~Bagrets, A.~Altland, and A.~Kamenev, {\it {Sachdev–Ye–Kitaev model as
  Liouville quantum mechanics}},  {\em Nucl. Phys.} {\bf B911} (2016) 191--205,
  [\href{http://xxx.lanl.gov/abs/1607.0069}{{\tt arXiv:1607.0069}}].

\bibitem{kitaev2018soft}
A.~Kitaev and S.~J. Suh, {\it {The soft mode in the Sachdev-Ye-Kitaev model and
  its gravity dual}},  {\em JHEP} {\bf 05} (2018) 183,
  [\href{http://xxx.lanl.gov/abs/1711.0846}{{\tt arXiv:1711.0846}}].

\bibitem{almheiri2015bulk}
A.~Almheiri, X.~Dong, and D.~Harlow, {\it {Bulk Locality and Quantum Error
  Correction in AdS/CFT}},  {\em JHEP} {\bf 04} (2015) 163,
  [\href{http://xxx.lanl.gov/abs/1411.7041}{{\tt arXiv:1411.7041}}].

\bibitem{maldacena2018eternal}
J.~Maldacena and X.-L. Qi, {\it {Eternal traversable wormhole}},
  \href{http://xxx.lanl.gov/abs/1804.0049}{{\tt arXiv:1804.0049}}.

\bibitem{allen1986spinor}
B.~Allen and C.~L{\"u}tken, {\it Spinor two-point functions in maximally
  symmetric spaces},  {\em Communications in Mathematical Physics} {\bf 106}
  (1986), no.~2 201--210.

\bibitem{allen1986vector}
B.~Allen and T.~Jacobson, {\it Vector two-point functions in maximally
  symmetric spaces},  {\em Communications in Mathematical Physics} {\bf 103}
  (1986), no.~4 669--692.

\end{thebibliography}\endgroup
\bibliographystyle{jhep}

\appendix
\section{SYK size effective action}
\label{sec:syk-size-effective-action}
Here, we work out some details related to the SYK effective size action. First, we derive the size effective action. We begin by noting that \(e^{- \mu n_j} = e^{- \frac{\mu}{2} + \ln \cosh \frac{\mu}{2}} (1 + 2 \tanh \frac{\mu}{2} \chi^L_j (- i \chi^R_j))\), and then simply expand the definition
\begin{align}
(0 | &\rho^{1/2}_{\beta} \chi^L_j(u_2) e^{- \mu n} \chi^L_j(u_1) \rho^{1/2}_{\beta} | 0) \\
&= e^{- N(\frac{\mu}{2} - \ln \cosh \frac{\mu}{2})}
(0 | \rho_{\beta}^{1/2} \chi^L_j(u_2) 
\left[\prod_{k = 1} 1 + 2 \tanh \frac{\mu}{2} \chi^L_k (- i \chi^R_k ) \right]
\chi^L_j(u_1) \rho_{\beta}^{1/2} | 0 ) \\
&= e^{- N(\frac{\mu}{2} - \ln \cosh \frac{\mu}{2})}
\sum_{m = 0}^N
\sum_{\{k_1, \ldots, k_m\}}
(-2 \tanh \frac{\mu}{2})^m
(0 | \rho_{\beta}^{1/2} \chi^L_j(u_2) 
\chi^L_{k_1} \cdots \chi^L_{k_m}
\chi^L_j(u_1) \rho_{\beta}^{1/2} \chi^L_{k_m} \cdots \chi^L_{k_1} | 0 ) \\
\label{eq:effective-size-action-path-integral-switch}
\begin{split}
&= e^{- N(\frac{\mu}{2} - \ln \cosh \frac{\mu}{2})}
\sum_{m = 0}^N
\sum_{\{k_1, \ldots, k_m\}}
(-2 \tanh \frac{\mu}{2})^m \\
&\qquad \langle \mathcal{T} \left\{
\chi_j(\frac{\beta}{2} + \epsilon + i u_2)
\chi_{k_1}(\frac{\beta}{2}) \cdots \chi_{k_m}(\frac{\beta}{2})
\chi_j(\frac{\beta}{2} - \epsilon + i u_1)
\chi_{k_m}(0) \cdots \chi_{k_1}(0)
\right\} \rangle_{\beta}
\end{split} \\
\begin{split}
&= e^{- N(\frac{\mu}{2} - \ln \cosh \frac{\mu}{2})}
\sum_{m = 0}^N
\sum_{\{k_1, \ldots, k_m\}}
(2 \tanh \frac{\mu}{2})^m \\
&\qquad \langle \mathcal{T} \left\{
\chi_j(\frac{\beta}{2} + \epsilon + i u_2)
(\chi_{k_1}(\frac{\beta}{2}) \chi_{k_1}(0)) \cdots (\chi_{k_m}(\frac{\beta}{2}) \chi_{k_m}(0))
\chi_j(\frac{\beta}{2} - \epsilon + i u_1)
\right\} \rangle_{\beta}
\end{split} \\
\label{eq:effective-size-action-G-replacement}
&= e^{- N(\frac{\mu}{2} - \ln \cosh \frac{\mu}{2})}
\sum_{m = 0}^N
\frac{(N 2 \tanh \frac{\mu}{2})^m}{m!}
\langle \mathcal{T} \left\{
\chi_j(\frac{\beta}{2} + \epsilon + i u_2)
G(\frac{\beta}{2}, 0)^m
\chi_j(\frac{\beta}{2} - \epsilon + i u_1)
\right\} \rangle_{\beta} \\
&= \langle \mathcal{T} \left\{ e^{- N S_{\mu}} 
\chi_j(\frac{\beta}{2} + \epsilon + i u_2) 
\chi_j(\frac{\beta}{2} - \epsilon + i u_1)
\right\} \rangle_{\beta},
\end{align}
where $S_\mu$ in the last line is defined in Eq. (\ref{eq:Smu}). In \eqref{eq:effective-size-action-path-integral-switch} we introduce a time-ordered path integral, whence all the fermions \(\chi\) become Grassmanian variables squaring to zero, and in \eqref{eq:effective-size-action-G-replacement} we use this property of Grassmanians to introduce \(G\). The denominator in the effective size expression is derived with similar manipulations.

Next, we find the saddle point for small \(\mu\). As discussed in the main text, we really just need to find the dependence of \(\lambda\) on \(\mu\). We note that
\begin{align*}
- X_1 \cdot X_2 &= \frac{1}{2}(X_2 - X_1)^2 + 1 =  \cosh \delta(X_1, X_2) \\
&= 
\coth \rho(\phi_1) \coth \rho(\phi_2) 
\sqrt{1 - \left(\frac{2 \pi}{L} \rho'(\phi_1)\right)^2} 
\sqrt{1 - \left(\frac{2 \pi}{L} \rho'(\phi_2) \right)^2} \\
&\qquad \left[ \frac{2}{\theta'(\phi_1) \theta'(\phi_2)} \left( \frac{L}{2 \pi} \sin \frac{\theta(\phi_1) - \theta(\phi_2)}{2} \right)^2 \right] \\
&\quad + \cosh (\rho(\phi_1) - \rho(\phi_2)) \cos (\theta(\phi_1) - \theta(\phi_2))
\end{align*}
where \(\delta\) is the geodesic distance. Thus, we can approximate
\begin{equation}
  \label{eq:g-as-h2-distance}
  [\theta'(\phi_1) \theta'(\phi_2)]^{\Delta} 
  G(\theta(\phi_2), \theta(\phi_1)) 
  \approx 
  c_{\Delta}
  [\cosh \delta(X_1, X_2) - 1]^{-\Delta} \sgn(\theta(\phi_2) - \theta(\phi_1))
\end{equation}
in the large \(L\) limit. Thus, we should find the opening angle \(\lambda(\mu)\) that minimizes
\begin{equation*}
- A - 2 \tanh \frac{\mu}{2} c_{\Delta} \cosh \delta(X_{\pi}, X_0)^{-\Delta}.
\end{equation*}

The basic quantities in the action are easiest to find by first considering the circle to have center at \(\rho = 0\) in Rindler coordinates, with radius \(r\) and segment angle \(\lambda\). The distance between the endpoints of the circular segment is simplest to find by the inner product in embedding coordinates,
\begin{equation*}
\cosh^2 r (1 - \tanh^2 r \cos \lambda).
\end{equation*}
The area of a single segment is given by a fraction \(\lambda\) of the area of the circle, plus the area of the triangular wedge, which we find from the interior angles (we call the one that is not \(2 \pi - \lambda\), \(\gamma\)) after another application of Gauss-Bonnet,
\begin{align*}
\frac{A}{2} &= \lambda \cosh r - 2 \gamma - \pi \\
\tan \gamma &= \sech r \tan \frac{\lambda - \pi}{2}.
\end{align*}
Using these expressions, we can expand the derivative of the action to leading order in \(L\) to find an equation for \(\lambda = \pi + \delta \lambda\),
\begin{multline*}
\left( \frac{2(\pi + \delta \lambda)}{L} \right)^{1 - 2 \Delta}
\left( \tan^2 \frac{\delta \lambda}{2} + \sec^2 \frac{\delta \lambda}{2} \frac{\sin \delta \lambda}{\pi + \delta \lambda} \right) \\
= 2 \tanh \frac{\mu}{2} \frac{\Delta c_{\Delta}}{2^{\Delta + 1}} \cos^{- 2 \Delta} \frac{\delta \lambda}{2}
\left( \tan \frac{\delta \lambda}{2} + \frac{2}{\pi + \delta \lambda} \right) + O(L^{-2}).
\end{multline*}
If we further expand to second order in \(\delta \lambda\), we find
\begin{equation*}
\delta \lambda = 2 \tanh \frac{\mu}{2} \frac{\Delta c_{\Delta}}{2^{\Delta}} \left( \frac{L}{2\pi} \right)^{1 - 2 \Delta}
- \left( 2 \tanh \frac{\mu}{2} \right)^2 \frac{1 - 2 \Delta}{\pi} \left( \frac{\Delta c_{\Delta}}{2^{\Delta}} \right)^2
\left( \frac{L}{2\pi} \right)^{2 - 4 \Delta}
+ O((2 \tanh \frac{\mu}{2} L^{1 - 2 \Delta})^{3}).
\end{equation*}
In general, for small \(\mu\) we can solve the equation for \(\delta \lambda\) order-by-order in a power series in \(2 \tanh \frac{\mu}{2} L^{1 - 2 \Delta}\). The behaviour of size can be understood as the response under small changes in the angle \(\lambda\), with the complication that we need to multiply by the appropriate derivative of \(\lambda\) over \(\mu\). 

For concrete computations using the geometric saddle point solution, we need to map from boundary time to location on the bulk curve, using that the former is an affine parameter for the latter. To this end, it is useful to transform between coordinate systems where one of the $\mu \ne 0$ segments is centered, as in the second panel of Figure~\ref{fig:SYK-size-saddle-point}, and a coordinate that is symmetric between the two segments, as in the third panel of Figure~\ref{fig:SYK-size-saddle-point}. This transformation is given by some $\ads_2$ isometry, which is simplest to express in the embedding coordinate. Suppose we start in the coordinate where the first segment is centered, as in the second panel of Figure~\ref{fig:SYK-size-saddle-point}. Then, the endpoints of the first segment are located at $(\cosh \rho, - \sinh \rho \sin \frac{\lambda - \pi}{2}, \pm \sinh \rho \cos \frac{\lambda - \pi}{2})$. The isometry that brings these points to points equidistant from the origin with $\theta = 0, \pi$ is the boost with parameter $\tanh \beta = \tanh \rho \sin \frac{\lambda - \pi}{2}$ generated by $E^{(E)}$. Explicitly, it is the embedding coordinate matrix
\begin{equation}
  \label{eq:upper-centered-to-symmetric-boost}
  F_{\beta} = \begin{pmatrix}
    \cosh \beta & \sinh \beta & 0 \\
    \sinh \beta & \cosh \beta & 0 \\
    0 & 0 & 1
  \end{pmatrix}.
\end{equation}
The inverse boost gives the transformation to the symmetric coordinate from the coordinate where the second segment is centered. In this way, we can always work in centered coordinates for the appropriate segment to map boundary times to bulk points.

Using this transformation, we compute the location of the bulk points corresponding to $\phi_{\pm} = \pi \pm \epsilon + i t_{\pm}$ for the saddle point solution corresponding to angle $\lambda$. Start with the $-$ coordinate, so use a coordinate system where the upper segment is centered as in the second panel of Figure~\ref{fig:SYK-size-saddle-point}. The radial coordinate $\rho$ is fixed by the requirement $\sinh \rho = L / 2 \lambda$. The angular coordinate is given by the affine parameter condition,
\begin{equation}
  \theta_{1-} = \frac{\lambda \phi_{-}}{\pi} - \frac{\lambda - \pi}{2}.
\end{equation}
Forming this into a coordinate $\tilde{X}_{<} = (\cosh \rho, \sinh \rho \sin \theta_{1-}, \sinh \rho \sin \theta_{1-})$, the coordinate in the symmetric system is given by $X_{<}(\lambda) = F_{\beta} \tilde{X}_{<}$. For the $+$ coordinate, we have
\begin{equation}
  \theta_{1+} = \frac{\lambda - \pi}{2} - \frac{\lambda}{\pi} (2 \pi - \phi_{+}),
\end{equation}
$\tilde{X}_{>} = (\cosh \rho, \sinh \rho \sin \theta_{1+}, \sinh \rho \cos \theta_{1+})$ in the coordinate where the second segment is centered, and $X_{>}(\lambda) = F_{- \beta} \tilde{X}_{>}$ in the symmetric system. To compute derivatives over $\lambda$, we use the definitions of $\theta_{\pm}$, $\rho$, $\beta$, and the useful identities
\begin{equation}
  \partial_{\theta_{\pm}} \tilde{X}_{\gtrless} = B^{(E)} \tilde{X}_{\gtrless}
  , \; \partial_{\rho} \tilde{X}_{\gtrless} = (\sin \theta_{\pm} E^{(E)} + \cos \theta_{\pm} P^{(E)}) \tilde{X}_{\gtrless}.
\end{equation}

\section{\(\ads\) space coordinates and symmetries}
\label{sec:ads-space-conventions}
For convenience, we collect here some \(\ads_d\), and in particular \(\ads_2\), coordinate systems and related expressions.
\subsection{Embedding coordinates}
\label{sec:embedding-coordinates}
A convenient definition of \(\ads_{d+1}\) involves the hyperboloid \(X^2 = -1\) in the space \(\mathbb{R}^{d + 2}\), with metric \(\eta\) of signature \((-, -, +, \ldots, +)\). We will also refer to the two timelike coordinates as \(T^0\) and \(T^1\), and in general start our numbering of embedding coordinates from \(0\). The global \(\ads_{d+1}\) space is defined as the universal covering of this hyperboloid, but we will also be interested in coordinate patches that cover only part of the global space.

The Killing vectors in \(\ads_{d+1}\) are the suitably restricted Killing vectors of the Lorentz group in the embedding space,
\begin{equation*}
K_{\mu\nu} = X_{\mu} \partial_{\nu} - X_{\nu} \partial_{\mu}.
\end{equation*}
It will be useful in what follows to identify a particular set of ``light-cone'' coordinates,
\begin{align*}
U^{\pm} &= X^2 \pm T^1 \\
V^{\pm} &= X^2 \pm T^0
\end{align*}
and to identify the Casimir
\begin{equation*}
C = \frac{1}{2} \sum_{j_1, j_2, k_1, k_2 = 1}^{p + q} \eta^{j_1 j_2} \eta^{k_1 k_2} K_{j_1 k_1} K_{j_2 k_2}.
\end{equation*}
\subsubsection{\(\ads_2\) Rindler coordinates}
\label{sec:ads2-rindler-coordinates}
The Rindler coordinate takes some boost, say \(K_{20}\), to be time translation. Orbits of this boost occur at the intersection of the constant \(T^{1}\) planes with the hyperboloid; an explicit coordinate choice is
\begin{align*}
T^1 &= \cosh \rho \\
V^{\pm} &= e^{\pm t_R} \sinh \rho
\end{align*}
with metric
\begin{equation*}
ds^2 = - \sinh^2 \rho dt_R^2 + d\rho^2.
\end{equation*}
Then the restrictions of the symmetry generators become
\begin{align*}
K_{20} &= \partial_{t_R} \\
K_{01} \pm K_{21} &= e^{\mp t_R} (\coth \rho \partial_{t_R} \pm \partial_{\rho}).
\end{align*}
\subsubsection{\(\ads_d\) Poincare coordinates}
\label{sec:ads-d-poincare-coordinates}
In the Poincare coordinate, some boost, say \(K_{21}\), becomes the naive coordinate ``dilatation'' when restricted to the projective boundary of the hyperboloid. An explicit coordinate system is
\begin{align*}
U^{+} &= \frac{1}{z} \\
X^j &= \frac{x^j}{z} \text{ for } j \not \in \{1,2\} \\
ds^2 &= \frac{dx^j dx_j + dz^2}{z^2}
\end{align*}
where indices are lowered on the \(x^j\) with the same signature as the \(X^j\). The restricted symmetry generators are
\begin{align*}
K_{21} &= - z \partial_z - x^j \partial_j \\
K_{2j} - K_{1j} &= \partial_j \text{ for } j \not \in \{1,2\} \\
K_{jk} &= x_j \partial_k - x_k \partial_j \text{ for } j \not \in \{1,2\} \\
K_{2j} + K_{1j} &= 2 x_j (x^k \partial_k + z \partial_z) - (x^2 + z^2) \partial_j.
\end{align*}
\subsubsection{\(\ads_d\) Global coordinates}
\label{sec:ads-d-global-coordinates}
In the ``global'' coordinate system, we choose the \(T^0-T^1\) rotation to give the local time translations. The explicit coordinates are
\begin{align*}
T^1 &= \sec \sigma \cos \tau, T^0 = \sec \sigma \sin \tau \\
X^j &= \tan \sigma \Omega_{d - 1}^j, \text{ for } j \not \in \{0 ,1\} \\
ds^2 &= \frac{- d\tau^2 + d\sigma^2 + \sin^2 \sigma d\Omega_{d - 1}}{\cos^2 \sigma},
\end{align*}
and the symmetry generators become
\begin{align*}
K_{10} &= - \partial_{\tau} \\
K_{jk} &\to \text{rotations of } \Omega_{d - 1} \text{ for } j, k \not \in \{0, 1\} \\
u^{\pm} &= \sigma \pm \tau \\
K_{1j} \pm i K_{0j} &= - e^{i u^{\pm}} x_j \partial_{\pm} - e^{- i u^{\mp}} x_j \partial_{\mp} + e^{\pm i \tau} (\csc \sigma \delta^k_j + x^k x_j) \partial_k.
\end{align*}
\section{Position space fermion two-point function}
\label{sec:bulk-fermion-2pt}
In this section, we extend the work \cite{allen1986spinor,allen1986vector} on geometric expressions for propagators in symmetric spaces to spinor representations in arbitrary dimension \(D = d + 1\), and curvature normalized to \(s_R = \frac{R}{D(D - 1)} \in \{0, 1, -1\}\). Call the geodesic distance from \(x\) to \(x'\), \(\delta(x, x')\), \(g_{\mu \nu}\) the metric tensor, and \(\Pi(x, x')_{\mu\nu'}\) the operator that parallel transports vectors along the shortest geodesic from \(x\) to \(x'\). We will repeat the convention established in \cite{allen1986vector} that primed (unprimed) indices correspond to indices that refer to the tangent space at \(x'\) (\(x\)), and omit the arguments \(x, x'\) where there is no ambiguity. The tangent vectors at the ends of the geodesic connecting \(x\) and \(x'\) are \(n_{\mu} = \nabla_{\mu} \delta\) (\(n^{(L)}\) without indices), and \(n_{\mu'} = \nabla_{\mu'} \delta\) (\(n^{(R)}\) without indices). Define also \(s_{x x'} = n_{\mu} n^{\mu} = n_{\mu'} n^{\mu'}\), which is \(1\) (\(-1\)) for points that are spacelike (timelike) separated. Then it can be shown \cite{allen1986vector} that any tensor acting on the tangent spaces at \(x\) and \(x'\) that is invariant (i.e. has zero Lie derivative) under the flow by an isometry can be written as scalar functions of the geodesic distance multiplying tensor products of \(\Pi\), \(g\), \(n^{(L)}\) and \(n^{(R)}\). We call such tensors ``invariant''. For example,
\begin{align}
p_{\mu\nu} &= g_{\mu\nu} - s_{x x'} n_{\mu} n_{\nu} \\
\label{eq:nabla-n-form}
\nabla_{\mu} n_{\nu} &= A(\delta) p_{\mu\nu}, \; 
\overline{A} = s_{x x'} A \\
\label{eq:nabla-nprime-form}
(\nabla_{\mu} n_{\nu'}) \Pi^{\nu'}_{\nu} &= B(\delta) p_{\mu\nu}, \;
\overline{B} = s_{x x'} B \\
\label{eq:nabla-pi-form}
(\nabla_{\mu} \Pi_{\nu \sigma'}) \Pi^{\sigma'}{}_{\sigma}
&= C(\delta) (p_{\mu\nu} n_{\sigma} - p_{\mu \sigma} n_{\nu})
= C(\delta) (g_{\mu\nu} n_{\sigma} - g_{\mu\sigma} n_{\nu})
\end{align}
where we have used \(\nabla_{n^{(L)}} n^{(L)} = 0\), \(\Pi n^{(R)} = - n^{(L)}\) (parallel transport of geodesic tangent vector), \(n^{\mu'} \nabla_{\nu} n_{\mu'} = 0\), \(\nabla_{n^{(L)}} \Pi(x, x') = 0\), and that parallel transport preserves all inner products to fix the forms of the above tensors. Note that we must have \(A + B \to 0\) as \(\delta \to 0\) since the components \(n^{(L)}\) approach \(-n^{(R)}\) (equivalently, \(\Pi \to 1\)). To find \(C\), use
\begin{align*}
s_{xx'} B p_{\mu\nu} &= - \nabla_{\mu} n_{\nu} - n^{\nu'} \nabla_{\mu} \Pi_{\nu' \nu} \\
&= - \nabla_{\mu} n_{\nu} + n^{\sigma} C (p_{\mu\nu} n_{\sigma} - p_{\mu \sigma} n_{\nu}) \\
&\implies C = \overline{A} + \overline{B}.
\end{align*}
Taking the trace of \eqref{eq:nabla-n-form} shows \(A = \frac{1}{d} \laplacian \delta\). Furthermore, we can find the derivatives of \(A\), \(B\) in several ways by taking derivatives along geodesics. For example,
\begin{align*}
\overline{A}' p_{\mu\nu}
&= n^{\sigma} \nabla_{\sigma} (\nabla_{\mu} n_{\nu})
= n^{\sigma} \nabla_{\mu} \nabla_{\sigma} n_{\nu} + R_{\nu\lambda \sigma \mu} n^{\sigma} n^{\lambda} \\
&= -(\nabla_{\mu} n^{\sigma}) (\nabla_{\sigma} n_{\nu}) - s_R s_{x x'} p_{\mu\nu}
= - (\overline{A}^2 + s_R s_{x x'}) p_{\mu\nu}.
\end{align*}
Likewise, by considering \(n^{\sigma} \nabla_{\sigma} (\nabla_{\mu} n_{\nu'})\) and \(n^{\sigma'} \nabla_{\sigma'} (\nabla_{\mu} n_{\nu})\), we find the relations
\begin{align*}
\overline{A}' &= - \overline{B}^2 = - (\overline{A}^2 + s_R s_{x x'}) \\
\overline{B}' &= - \overline{A} \overline{B}.
\end{align*}
Finally, since all fields arise from the same principal bundle (so only the generators change in the ``spin connection''), the parallel transport operator \(S(x, x')\) for any associated bundle satisfies
\begin{equation*}
\nabla_{\mu} S(x, x') = C g_{\mu \nu} n_{\sigma} \Sigma^{\nu \sigma} S(x, x')
\end{equation*}
where \(\Sigma^{\nu\sigma}\) are the appropriate spin group generators.

Now we specialize to the case of spinors. Fix some vielbien by a choice of an appropriate vector-valued one-form \(\sigma\). Then if we define \(\Gamma_{\mu} = \gamma_a \sigma^a{}_{\mu}\) and \(\slashed{n} = \Gamma_{\mu} n^{\mu}\), we have
\begin{equation*}
\nabla_{\mu} S(x, x') = \frac{C}{4} [\Gamma_{\mu}, \slashed{n}] S(x, x').
\end{equation*}
For a different choice of vielbien \(\sigma(x) = \tilde{\sigma}(x) \Lambda(x)\), \(\tilde{S}(x, x') = \Lambda(x) S(x, x') \Lambda(x')^{-1}\). Next, we use that there are no invariant totally antisymmetric tensors with more than 1 index (of course, we take all these indices to be at a single point, say \(x\)). This can be shown by induction, or simply by noting all invariant tensors with indices at a single point \(x\) must be built out of sums of products of \(g\) and \(n^{(L)}\)\footnote{We need to consider discrete reflections to eliminate the totally antisymmetric tensor.}. Call the invariant two-point function of spinors \(G_{\psi}(x, x')\). Since traces of \(G_{\psi}(x, x') S(x, x')^{-1}\) against products of the \(\Gamma_{\mu}\) give invariant tensors with indices at a single point, the most general form of the fermion two-point function is
\begin{equation*}
G_{\psi}(x, x') = (F_0(\delta) + F_1(\delta) \slashed{n}) S(x, x')
\end{equation*}
for scalar functions of the geodesic distance \(F_0\), \(F_1\). It will also be useful to have the covariant derivative and Dirac operator
\begin{align}
\label{eq:g-psi-covariant-derivative}
\nabla_{\mu} G_{\psi}(x, x') &= 
((F_0'(\delta) + F_1'(\delta) \slashed{n}) n_{\mu} + A F_1(\delta) p_{\mu\nu} \Gamma^{\nu}) S(x, x')
+ G_{\psi}(x, x') S(x, x')^{-1} \frac{C}{4} [\Gamma_{\mu}, \slashed{n}] S(x, x') \\
\slashed{\nabla} G_{\psi}(x, x') &= \Gamma^{\mu} \nabla_{\mu} G_{\psi}(x, x')
= \left[ (F_0'(\delta) + \frac{d}{2} C F_0(\delta)) \slashed{n} + 
s_{x x'} (F_1'(\delta) + \frac{d}{2} (\overline{A} - \overline{B}) F_1(\delta)) \right] S(x, x').
\end{align}
The Dirac equation when \(x \ne x'\), \((\slashed{\nabla} - m) G_{\psi}(x, x') = 0\), reduces to the pair of equations
\begin{align*}
m F_1(\delta) &= F_0'(\delta) + \frac{d}{2} C F_0 \\
F_1'(\delta) + &\frac{d}{2}(\overline{A} - \overline{B}) F_1(\delta) - s_{x, x'} m F_0 = 0.
\end{align*}
In the non-flat case \(s_R \ne 0\), there is a useful function \(K = - A / B\), whence from our above relations we find \(K''(\delta) = - s_R s_{x x'} K\) and \(\laplacian K = - s_R D K\), so \(A = - s_R K / K'\). Then in terms of \(y = \frac{1 + K}{2}\) we can write a second-order equation for \(F_0\),
\begin{equation}
\label{eq:F0-2ndorder-diffeq}
\left[ y(y - 1) \partial_y^2 + D (y - \frac{1}{2}) \partial_y + ((\frac{d}{2})^2 + s_R m^2 + \frac{d}{4 y}) \right] F_0 = 0.
\end{equation}
From here, we consider \(s_R = -1\). Then the solution of \eqref{eq:F0-2ndorder-diffeq} that is properly normalized to a delta-function singularity in Euclidean signature and has the right decay at infinity for the Poincare vacuum is
\begin{align*}
K(\delta) &=
\begin{cases}
\cosh \delta & s_{xx'} > 0 \\
\cos \delta & s_{xx'} < 0
\end{cases}, \, y = \frac{1 + K}{2} \\
N_{\Delta, d} &= \frac{1}{\Omega_d}
\frac{\Gamma(\Delta + \frac{1}{2}) \Gamma(\Delta - \frac{d}{2})}{2^d \Gamma(2(\Delta - \frac{d}{2}) + 1) \Gamma(\frac{d+1}{2})}, \, \Omega_d = \frac{2 \pi^{\frac{d+1}{2}}}{\Gamma(\frac{d+1}{2})} \\
F_0(\delta) &= N_{\Delta, d} m f_0(y), \, F_1(\delta) = N_{\Delta, d} K'(\delta) f_1(y) \\
f_0(y) &= y^{-\Delta} \pFq{2}{1}(\Delta + \frac{1}{2}, \Delta - \frac{d}{2}; 2 (\Delta - \frac{d}{2}) + 1; \frac{1}{y}) \\
f_1(y) &= - \frac{\Delta - \frac{d}{2}}{2} y^{- (\Delta + 1)} \pFq{2}{1}(\Delta + \frac{1}{2}, \Delta - \frac{d-2}{2}; 2 (\Delta - \frac{d}{2}) + 1; \frac{1}{y})
\end{align*}
where as long as \(\Delta > 0\), we can choose \(\Delta = \frac{d}{2} \pm m\). We fix the branches of \(\pFq{2}{1}\) for the Lorentzian Wightmann function by analytic continuation of a Euclidean time coordinate \(\tau \to \epsilon + i t\). This is well-defined (meaning all relevant covariant derivatives continue in a consistent way) as long as we continue with respect to a Euclidean time \(\tau\) such that \(\partial_{\tau}\) is a Killing vector orthogonal to a family of hypersurfaces (in other words, the metric can be taken independent of \(\tau\), and with no cross-terms involving \(d \tau\)). 

The ``expectations'' of symmetry generators, i.e. expressions like \(\langle \psi(x) E \overline{\psi}(x') \rangle\), are especially simple to compute in this formalism. Given the vector field \(V\) generating the isometry of the manifold associated to \(E\), we have
\begin{equation*}
\langle \psi(x) E \overline{\psi}(x') \rangle
= i \nabla_V G_{\psi}(x, x'),
\end{equation*}
where the covariant derivative acts on the unprimed coordinate. We also note that for practical computations, \(S(x, x')\) can be found explicitly as the spinor transformation, smoothly connected to the identity, corresponding to \(\Lambda_{ab'}(x, x') = \sigma^{\mu}_a(x) \Pi_{\mu\nu'} \sigma^{\nu'}_{b'}(x')\). \(\Pi_{\mu\nu'}\) itself can be found from \(\nabla_{\mu} n_{\nu'}\).

\subsection{\(\ads_2\) propagator and generator expectation values}
\label{sec:ads2-prop-and-expectations}
Our main focus is on the Rindler coordinate on \(\ads_2\). This is the coordinate with Euclidean metric
\begin{equation*}
ds^2 = \sinh^2 \rho d\tau^2 + d\rho^2.
\end{equation*}
We take the natural vielbien \(\sigma(x) = \sinh \rho e_0 d\tau + e_1 d\rho\).  The bulk points we consider in the main text are at the same \(\rho\) coordinate, but different \(\tau\). We consider imaginary point splitting by some fixed amount \(\epsilon\). In fact, for different \(\rho\) this corresponds to different regimes of geodesic distance. 

If we first take the \(\rho \to \infty\) limit, then we should consider the propagator in the limit of \emph{large} geodesic distance. In this limit, \(\slashed{n} \to \gamma^1\). It remains to find \(S(x, x')\). If we take a vielbien \(\tilde{\sigma}(x, x')^a{}_{\mu} = \Pi_{\mu \nu'} \sigma^{a \nu'}(x')\), then \(\tilde{S}(x, x') = 1\), since we have chosen a non-coordinate basis in which parallel transport along the geodesic from \(x\) to \(x'\) is trivial. In the original \(\sigma\) vielbien, \(S(x, x') = \Lambda\) the spinor Lorentz transformation, smoothly connected to the identity, that corresponds to \(\Pi_{ab}(x, x') = \sigma(x)_a^{\nu} \sigma(x)_{b}^{\nu'} \Pi_{\nu \nu'}\). If we take \(\rho \to \infty\) at fixed \(\epsilon\), then \(\Lambda\), the Lorentz transformation taking \(n^{(R)} \to - n^{(L)}\) in the non-coordinate basis given by \(\sigma\), becomes a rotation by (minus) \(\pi\), so \(S(x, x') \to e^{- \pi \Sigma^{01}} = - \gamma^0 \gamma^1\). Then the large-\(\rho\) limit becomes
\begin{equation}
\label{eq:large-rho-bulk-correlator-limit}
G_{\psi}(x, x') \xrightarrow[\rho \to \infty]{}
\sech^{2 \Delta} \rho (2 N_{\Delta, 1} |m|) \left( \sin^2 \frac{\tau - \tau'}{2} \right)^{- \Delta}
\left( \frac{1 - \sgn \left( \frac{\Delta - \frac{d}{2}}{m} \right) \gamma^1}{2} \right)
\gamma^0
,
\end{equation}
where the last term is a projector onto a certain eigenspace of \(\gamma^1\). This is an indication that there is only a single fermion component on the boundary.

In the small geodesic distance limit, \(F_0(\delta) \to - m G_{d}(\delta)\), where \(G_{d}\) is the Green function for the flat Laplacian in dimension \(d + 1\). We also have that \(F_1(\delta) \to - G_d'(\delta)\). The vector \(n \to \csch \rho \partial_{\tau}\), and \(S(x, x') \to 1\).

The main use of this propagator in this paper is to compute the expectation of the generators \(E - B\); call the vector generating this isometry \(V\). Continuing the equation \(i \nabla_V G_{\psi}\) to Euclidean signature, we find that we need to compute
\begin{equation*}
- \nabla_{V^{(E)}} G_{\psi}(x, x'), \;
V^{(E)} = \sin \tau \partial_{\rho} + (\coth \rho \cos \tau - 1) \partial_{\tau}.
\end{equation*}
The exact expression is then given by taking the inner product of \(V^{(E)}\) with \eqref{eq:g-psi-covariant-derivative} and analytically continuing. For this it is useful to have the (Euclidean) spinor propagator
\begin{equation}
  \label{eq:same-rho-spinor-prop}
  S((\tau, \rho), (\tau', \rho)) = \frac{1 - \cosh \rho \tan \frac{\tau - \tau'}{2} \gamma^2}{\sqrt{1 + \cosh^2 \rho \tan^2 \frac{\tau - \tau'}{2}}}.
\end{equation}

\section{Fermion modes in \(\ads_{2}\)}
\label{sec:light-fermions}
In this section, we give fermion mode solutions corresponding to the natural time in several \(\ads_2\) coordinates. These modes serve four roles in this work. First, they give an explicit consistent quantization of the ``unusual'' fermions with boundary dimension \(\Delta < \frac{d}{2}\). Second, the Fourier transform of a reconstruction kernel can be read off of the modes. Third, they are used to show that our regularization of the kernel for \(\Delta < \frac{d}{2}\) is correct. Finally, they have been used to check the two-point function we derive by a mode sum. We illustrate these points in detail for the example of Poincare coordinates.

\subsection{Poincare coordinates}
\label{sec:poincare-coordinate-modes}

We start in the simpler Poincare coordinates to illustrate some general points. The metric is
\begin{equation}
ds_{d+1}^2 = \frac{dz^2 + \eta_{ab} dx^a dx^b}{z^2},
\end{equation}
with Dirac operator
\begin{equation}
\slashed{\nabla} = z \gamma^j \partial_j - \frac{d}{2} \gamma^z.
\end{equation}
We are looking for modes of the equation \((\slashed{\nabla} - m) \psi = 0\). The boundary fermion will have dimension \(\Delta = \frac{d}{2} \pm m\) depending on our particular mode choice, and we allow either sign as long as \(\Delta > 0\). To emphasize this point, when \(|m| < \frac{d}{2}\), there are \emph{two} consistent quantizations, one with \(\Delta > d / 2\), and one with \(\Delta < d / 2\). As we will show, both choices give rise to normalizable modes in \(\ads_D\). Define \(\pi^{g^z}\) to be the projector onto some eigenvalue of \(\gamma^z\), \(\pi^{g^z} = \frac{1}{2}(1 + g^z \gamma^z)\). Define also \(|p| = \sqrt{- p^a p^b \eta_{ab}}\) and \(\slashed{n} = i \gamma^a p^b \eta_{ab} / |p|\). The (matrix-valued) function
\begin{align}
\label{eq:poincare-bulk-mode-operator}
F_{g^z p}(x, z) &=
e^{i p \cdot x}
\sqrt{\frac{|p|}{2}} \frac{z^{\frac{d+1}{2}}}{(2 \pi)^{\frac{d-1}{2}}}
\left[
J_{\Delta - \frac{d + 1}{2}}(z |p|) + \slashed{n} \gamma^z J_{\Delta - \frac{d - 1}{2}}(z |p|)
\right]
\pi^{g^z} \\
\label{eq:poincare-bulk-mode-operator-scalar-derivative}
&= (\slashed{\nabla} + m + \gamma^z) \gamma^z \left[ \frac{e^{i p \cdot x}}{\sqrt{2 |p|}} \left( \frac{z}{2 \pi} \right)^{\frac{d-1}{2}} J_{\Delta - \frac{d-1}{2}}(z |p|) \right]
\pi^{g^z}
\end{align}
solves the Dirac equation with \(m = g^z (\Delta - \frac{d}{2})\). In terms of this function, the normalized modes of the Dirac equation (associated to the Poincare Killing vector) are
\begin{align}
\label{eq:poincare-fermion-bulk-modes}
\psi^j_{g^z p}(x, z) &= F_{g^z p}(x, z) u^j_{g^z}(p) \\
u_{g^z}^j(p) &= \Lambda_{1/2}^{-1}(p) u_{g^z}^j(0).
\end{align}
where \(u_{g^z}^j(0)\) is a basis for the \(\gamma^z = g^z\) eigenspace, \(\gamma^z u^j_{g^z}(0) = g^z u^j_{g^z}(0)\), and \(p\) is restricted to be timelike. \(\Lambda_{1/2}(p)\) is the Lorentz boost that takes the timelike vector \((\sgn p^0 \sqrt{-p^2}, 0, \ldots) \to p\). There are solutions with spacelike \(p\), but these are not normalizable in the bulk. These modes are normalized according to
\begin{equation}
\label{eq:poincare-modes-completeness-relation}
\int d^{d-1} x dz z^{-d} \psi_{g^z p}^j(x, z)^{\dagger} \psi_{g^z q}^j(x, z)
= \delta^{jk} \delta^d(p - q).
\end{equation}
As we take \(z \to 0\), the dominant behaviour is
\begin{equation}
F_{g^z p}(x, z) \xrightarrow[z \to 0]{}
z^{\Delta} \left(
\frac{e^{i p \cdot x}}{(2 \pi)^{\frac{d - 1}{2}} \Gamma(\Delta - \frac{d-1}{2})} \left( \frac{|p|}{2} \right)^{\Delta - \frac{d}{2}}
\right) \pi^{g^z}.
\end{equation}
Notice that this function is proportional to \(\pi^{g^z}\). This is how we get the correct (reduced) number of fermion components on the boundary.

We can construct smearing functions by the following procedure. First, we fix some notation. The bulk fermion field has the mode expansion
\begin{equation}
\label{eq:poincare-bulk-field-expansion}
\psi(z, x) = \int d^{d-1} p \int_{|\vec{p}|}^{\infty} dp^0 \sum_j \psi_{g^z p}^j(x, z) c_{p}^j + \psi_{g^z -p}^j(x, z) d_{p}^{j\dagger},
\end{equation}
while the boundary is
\begin{multline}
\chi_0(x) = 
\frac{1}{\Gamma(\Delta - \frac{d - 1}{2})(2\pi)^{\frac{d-1}{2}}}
\int d^{d-1} p \int_{|\vec{p}|}^{\infty} dp^0 
\sum_j
e^{i p \cdot x}
\left(\frac{|p|}{2}\right)^{\Delta - \frac{d}{2}}
u_{g^z}^j(p) c_{p}^j \\
+ 
e^{- i p \cdot x}
\left(\frac{|p|}{2}\right)^{\Delta - \frac{d}{2}}
u_{g^z}^j(-p) d_{p}^{j\dagger}
\end{multline}
Define \(\overline{u} = u^{\dagger}(-i \gamma^0)\). We then have
\begin{align}
\label{eq:poincare-boundary-to-mode-integrals}
c_p^j &= \frac{\Gamma(\Delta - \frac{d-1}{2})}{(2 \pi)^{\frac{d+1}{2}}} \left(\frac{|p|}{2}\right)^{\frac{d}{2} - \Delta} \overline{u_{g^z}^j(p)} \slashed{n}
\int d^d x e^{- i p \cdot x} \chi_0(x) \\
d_p^{j \dagger} &= \frac{\Gamma(\Delta - \frac{d-1}{2})}{(2 \pi)^{\frac{d+1}{2}}} \left(\frac{|p|}{2}\right)^{\frac{d}{2} - \Delta} \overline{u_{g^z}^j(- p)} \slashed{n}
\int d^d x e^{i p \cdot x} \chi_0(x);
\end{align}
note that here all momenta, including \(\slashed{n}\) in both expressions, have \(p^0 > 0\). Now, we re-insert these operators into \eqref{eq:poincare-bulk-field-expansion}. It is useful to have the formula
\begin{equation}
\sum_j u_s^j(p) \overline{u_s^j(p)} \slashed{n} = \sgn p^0 \pi^{g^z}.
\end{equation}
Then for
\begin{equation}
\label{eq:poincare-kernel-fourier-transform}
K_{\Delta}(x, z) = 
\frac{\Gamma(\Delta - \frac{d - 1}{2})}{(2 \pi)^{\frac{d + 1}{2}}}
\int d^d p \theta(- p^2) \left( \frac{|p|}{2} \right)^{\frac{d}{2} - \Delta}
F_{g^z p}(x, z)
\end{equation}
we find
\begin{equation*}
\psi(x, z) = \int d^d x' K_{\Delta}(x - x', z) \chi_0(x')
\end{equation*}
as an operator equation. 

A completely similar sequence of steps in other coordinate systems gives the Fourier transform of the reconstruction kernel directly from the bulk mode solutions. The key ingredient is that, near the boundary, the mode becomes proportional to the eigenspace of some \(\gamma\)-matrix. This particular type of decay is easiest to anticipate by examining the Dirac operator in a given coordinate system. A general feature is that the Fourier transform of the reconstruction kernel is given by the spinor operator that is
\begin{enumerate}
\item an eigenfunction of the flow by the Killing vector associated to the time coordinate,
\item takes arbitrary fixed spinors to a solution of the Dirac equation (this is \eqref{eq:poincare-bulk-mode-operator} in the Poincare coordinate),
\item and is normalized to approach \((z')^{\Delta} \pi e^{i p \cdot x}\) (up to numerical factors and functions of other coordinates for which there is no translation symmetry) for some projector \(\pi\) and some coordinate \(z'\) tending to zero as the conformal boundary is approached.
\end{enumerate}
This smearing function will correspond to approaching the conformal boundary at constant \(z'\).

We directly compute the position space smearing function for \(d = 1\). We use the integral
\begin{equation}
\label{eq:1d-poincare-smearing-integral}
\int_0^{\infty} d\mu \cos (\mu \frac{x^{0} - y^0}{z}) \mu^{-\Delta} J_{\Delta}(\mu)
= \begin{cases}
0 & z < |x^0 - y^0| \\
\frac{2^{-\Delta} \sqrt{\pi}}{\Gamma(\Delta + 1/2)}
(\frac{z^2 - (x^0 - y^0)^2}{z^2})^{\Delta - 1/2} & \text{otherwise};
\end{cases}
\end{equation}
and the form \eqref{eq:poincare-bulk-mode-operator-scalar-derivative} of the bulk mode operator, and find
\begin{align}
K_{\Delta}(t, z) 
\label{eq:poincare-mode-sum-kernel-result-derivative}
&= \frac{\Gamma(\Delta)}{2 \sqrt{\pi} \Gamma(\Delta + \frac{1}{2})}
z^{- \Delta} (\slashed{\nabla} + \frac{1}{2} g^z) g^z (z^2 - t^2)^{\Delta - \frac{1}{2}} \theta(z - |t|)\pi^{g^z} \\
\label{eq:poincare-kernel-from-mode-sum}
&= \frac{m \Gamma(\Delta)}{\sqrt{\pi} \Gamma(\Delta + \frac{1}{2})}
\frac{- t \gamma^0 + z \gamma^z}{\sqrt{-t^2 + z^2}} 
\left( \frac{-t^2 + z^2}{z} \right)^{\Delta - 1} \pi^{g^z} \theta(z - |t|),
\end{align}
where the second line is valid only for \(\Delta > 1/2\) because of derivatives of the step function. If the relevant integrals converge, we can just use this smearing function directly to construct bulk quantities. For \(\Delta < 1/2\), there is a divergence in this kernel on the light cone that is not integrable. This derivation assures us that the divergence is not spurious; it comes from unregulated arbitrarily high-momentum modes.

There are several possibilities to regulate the \(\Delta < 1/2\) divergence. One is to take the derivatives in \eqref{eq:poincare-mode-sum-kernel-result-derivative} after integrating against the kernel (since the Dirac operator is independent of time). The most immediate is to analytically continue in \(\Delta\). All that we need to reproduce the bulk modes is that the integral against boundary modes, \(\propto \int K_{\Delta}(t - u, z) e^{i \omega u}\), gives the properly normalized bulk modes, which themselves are analytic in \(\Delta\). If we define the integral of the kernel against analytic functions by taking a ``figure-eight''
contour around the poles at \(t = \pm z\) (and normalize by \(e^{i \pi \Delta} \cos \pi \Delta\)),%
\begin{tikzpicture}[pole/.style={fill=black, shape=circle, inner sep=0, minimum size=3pt},
  use Hobby shortcut,
  every path/.style={line width=1pt, white, double=black, double distance=0.5pt}]
  \node[pole, name={left pole}] at (0, 0) {};
  \node[pole, name={right pole}] at (1, 0) {};
  \draw
  ([closed]-0.2, 0) .. (0, 0.2) .. ([blank=soft]0.5, 0) .. ([blank=soft]1, -0.2) .. (1.2, 0) .. (1, 0.2) .. (0.5, 0) .. (0, -0.2);
  \draw[use previous Hobby path={invert soft blanks}];
  \path[
  black, line width=0.5pt,
  decoration={markings,
    mark=at position 0.2 with{\arrow{>[width=4pt,length=3pt]}},
    mark=at position 0.34 with{\arrow{>[width=4pt,length=3pt]}},
    mark=at position 0.7 with{\arrow{>[width=4pt,length=3pt]}},
    mark=at position 0.84 with{\arrow{>[width=4pt,length=3pt]}},
  }, postaction=decorate]
  ([closed]-0.2, 0) .. (0, 0.2) .. (0.5, 0) .. (1, -0.2) .. (1.2, 0) .. (1, 0.2) .. (0.5, 0) .. (0, -0.2);
\end{tikzpicture},%
then for \(\Delta > 1/2\) this gives the correct answer, and for \(\Delta < 1/2\) gives the correct bulk modes by analytic continuation.

The analytic continuation method is also related to a simple high-frequency regulator. This can be done by giving an exponential energy damping \(e^{- \epsilon |\omega|}\) on each mode in \eqref{eq:poincare-kernel-fourier-transform}. Instead of the sharp step in \eqref{eq:1d-poincare-smearing-integral}, the integral defining the regulated \(K_{\Delta}(t, z)\) becomes, calling \(\alpha = (x^0 - y^0)/z\),
\begin{align}
\label{eq:regulated-1d-poincare-kernel}
\Re\left[
\int_0^{\infty} d\mu e^{- (\epsilon + i \alpha) \mu} \mu^{-\Delta} J_{\Delta}(\mu)
\right]
&\to \frac{2^{-\Delta}}{\Gamma(\Delta+1)}\frac{1}{\sqrt{\alpha^2 + \epsilon^2}}
\Im \left( 
\pFq{2}{1}\left(\frac{1}{2}, 1; \Delta + 1; \frac{1}{\alpha^2 - i \epsilon}\right)\right) \\
&\to
\frac{\sqrt{\pi}}{2^{\Delta} \Gamma(\Delta + 1/2)} \frac{\Re \left( e^{-i \pi \Delta} (1 - (\alpha^2 - i \epsilon))^{\Delta - 1/2} \right)}{\cos (\pi \Delta)}.
\end{align}
We can now freely take derivatives of this integral as in \eqref{eq:poincare-mode-sum-kernel-result-derivative} to find the regulated Poincare kernel; the difference will only be non-vanishing in \(\epsilon\) near \(\alpha = 1\). Integrating this regulated kernel against analytic functions is the same as our contour prescription in the limit \(\epsilon \to 0\). Since we only consider integrals of the kernel against analytic functions, and the analyticity argument is simpler than carrying out explicit regulated integrals, in other coordinate systems we will simply use the analytic continuation as the definition of the kernel for \(\Delta < 1/2\).

\subsection{\(d=1\) global coordinates}
\label{sec:d1-global-coordinates}

In global coordinates, we have
\begin{equation*}
ds_{\ads_2} = \frac{- d\tau^2 + d\sigma^2}{\cos^2 \sigma}.
\end{equation*}
We choose vielbiens
\begin{equation*}
e_a = \cos \sigma \partial_a;
\end{equation*}
with this choice the nonzero component of the spin connection is
\begin{equation*}
w_{\tau 0 1} = - w_{\tau 1 0} = - \tan \sigma
\end{equation*}
and the Dirac operator is
\begin{equation*}
\slashed{\nabla} = \cos \sigma (\gamma^0 \partial_{\tau} + \gamma^1 \partial_{\sigma}) + \frac{\sin \sigma}{2} \gamma^1.
\end{equation*}
The normalized positive frequency solutions to this equation are given by
\begin{multline}
\label{eq:global-positive-modes}
\psi^{(+)}_n(\sigma, \tau) =
\frac{\sqrt{2 n! \Gamma(2 \Delta + n)}}{2^{\Delta} \Gamma(\Delta + n)}
e^{- i (\Delta + n) \tau} \cos^{\Delta}(\sigma) \\
\left(
\cos \left( \frac{\pi}{4} - \frac{\sigma}{2} \right)
P_n^{(\Delta - 1, \Delta)}(\sin \sigma)
+ i \gamma^2 \sin \left( \frac{\pi}{4} - \frac{\sigma}{2} \right)
P_n^{(\Delta, \Delta - 1)}(\sin \sigma)
\right) u_s
\end{multline}
where \(P_n^{(\alpha, \beta)}\) are the Jacobi polynomials, \(u_s\) are eigenvectors of \(\gamma_1\) with eigenvalue \(s\), such that \(\Delta = 1/2 - sm\), and \(\gamma^2 = \gamma^0 \gamma^1\). If we work in a basis where \(\gamma^2\) and \(u_s\) are real, then we can take the negative frequency modes just the complex conjugates of \eqref{eq:global-positive-modes}; in general they are
\begin{multline}
\label{eq:global-negative-modes}
\psi^{(-)}_n(\sigma, \tau) =
\frac{\sqrt{2 n! \Gamma(2 \Delta + n)}}{2^{\Delta} \Gamma(\Delta + n)}
e^{i (\Delta + n) \tau} \cos^{\Delta}(\sigma) \\
\left(
\cos \left( \frac{\pi}{4} - \frac{\sigma}{2} \right)
P_n^{(\Delta - 1, \Delta)}(\sin \sigma)
- i \gamma^2 \sin \left( \frac{\pi}{4} - \frac{\sigma}{2} \right)
P_n^{(\Delta, \Delta - 1)}(\sin \sigma)
\right) u_s.
\end{multline}
These modes are orthonormal under the inner product
\begin{equation*}
\int_{-\pi/2}^{\pi/2} d\sigma \sec \sigma \psi^s_n(\tau, \sigma)^{\dagger} \psi^{s'}_m(\tau, \sigma) = \delta_{nm} \delta_{s s'}.
\end{equation*}

\subsection{\(d=1\) Rindler coordinates}
\label{sec:d-1-rindler-coordinates}

It is convenient to make the change of variable to \(u = - \ln \tanh \frac{\rho}{2}\). The new coordinates are
\begin{align*}
T^1 &= \coth u = \cosh \rho \\
V^{\pm} &= \csch u e^{\pm t_R} = \sinh \rho e^{\pm t_R}.
\end{align*}
The metric in these coordinates becomes
\begin{equation*}
ds^2 = \frac{- dt_R^2 + du^2}{\sinh^2 u}.
\end{equation*}
The boundary is located at \(u \to 0\). The natural vielbiens in this coordinate are
\begin{equation*}
e_a = \sinh u \partial_a
\end{equation*}
and the Dirac operator is
\begin{equation*}
\slashed{\nabla} = \sinh u \slashed{\partial} - \frac{1}{2} \cosh u \gamma^1.
\end{equation*}
To find the modes, we solve the Dirac equation \((\slashed{\nabla} - m) \psi(t_R, \rho) = 0\). The normalized expressions are
\begin{align*}
\psi(t_R, u) &= N_{\omega} e^{- i \omega t_R} (-z)^{\tilde{c}/2} (1-z)^{\frac{1-\tilde{c}}{2}} F_{\omega, m}(z) \pi_s u_s \\
F_{\omega, m}(z) &= \tilde{c} \pFq{2}{1}(a, b; \tilde{c}; z)
- a \gamma^2 (-z)^{\tilde{c}} \pFq{2}{1}(\tilde{c} + a, \tilde{c} + b; 1+\tilde{c}; z) \\
z &= - \sinh^2 \frac{u}{2} \\
a &= i \omega, b = - i \omega, \tilde{c} = \frac{1}{2} + m \gamma^1 \\
N_{\omega} &= \frac{1}{\Delta}
\sqrt{\frac{2 \cosh (\pi \omega) B(\Delta + i \omega, \Delta - i \omega)}{\pi B(\Delta, \Delta)}}.
\end{align*}
A mode sum for the kernel for \(\Delta > 1/2\) gives \eqref{eq:large-delta-simple-rindler-kernel}.

\section{Details on the reconstruction kernel}
\label{sec:reconstruction-kernel-details}
A very direct way to derive the kernel is the mode sum approach. We have carried out the mode sum both in Poincare and Rindler coordinates for \(\Delta > 1/2\) to confirm our expressions. That approach also illustrates that the \(\Delta < 1/2\) case can be treated either by analytic continuation from \(\Delta > 1/2\), or equivalently by regulating high-momentum modes on the boundary.

To connect with the invariant description of reconstruction in \eqref{eq:coordinate-invariant-reconstruction}, and to show more explicitly the role that choice of coordinate and vielbien plays in the reconstruction kernel, we write the kernel using the geometric quantities described in \ref{sec:bulk-fermion-2pt}. First, call \(N_{\mu}\) the normal vector the conformal boundary, and \(\slashed{N} = \Gamma_{\mu} N^{\mu}\). Note that \(\slashed{N}\) depends on how we approach the boundary, which is different in different coordinate systems. Now solve the Dirac equation for spacelike separation using the same ansatz (and notation) as in \ref{sec:bulk-fermion-2pt}, but now demanding regularity as points approach each other. The solution is
\begin{align*}
G_K(x, x') &= (F_0(\delta) + F_1(\delta) \slashed{n}) S(x, x') \\
F_0 &= m y^{-\frac{d}{2}} \pFq{2}{1}(\Delta - \frac{d}{2}, d - \Delta - \frac{d}{2}; \frac{d+1}{2}; 1-y) \\
F_1 &= \frac{2 \sqrt{y - 1} (\Delta - \frac{d}{2})^2 \pFq{2}{1}(\Delta + \frac{1}{2}, d - \Delta + \frac{1}{2}; \frac{d+3}{2}; 1 - y)}{d + 1}.
\end{align*}
It can be checked that the kernel in either coordinate system can be written (using \(z'\) as a general coordinate tending to zero at the conformal boundary, for example \(\sech \rho\) in Rindler)
\begin{align*}
K_{\Delta}(x, x') &\propto 
\lim_{z' \to \infty} \sqrt{- g(x')} (z')^{\Delta} G_K(x, x') \slashed{N} \\
&\propto \lim_{z' \to \infty} \sqrt{- g(x')} (z')^{\Delta}
S(x, x') y^{\Delta - d} (1 - \frac{\Delta - \frac{d}{2}}{m} n^{(R)}_{\mu'} \Gamma^{\mu'}(x')) \slashed{N}.
\end{align*}
The form of the reconstruction kernel in Rindler and Poincare coordinate systems, \eqref{eq:large-delta-simple-rindler-kernel} and \eqref{eq:poincare-kernel-from-mode-sum} respectively, has been chosen to reflect the second equality here. The matrix appearing in the front of both kernels is \(S(x, x')\),
\begin{equation*}
\lim_{z' \to 0} S(x, x') \propto
\begin{cases}
\frac{\gamma^a x_a + \gamma^z z}{\sqrt{x_a x^a + z^2}} \gamma^z & \text{Poincare} \\
\frac{1 + e^{\rho} \tanh (t_R/2) \gamma^2}{\sqrt{1-e^{2 \rho} \tanh^2 \left(\frac{t_R}{2} \right)}} &
\text{Rindler}
\end{cases},
\end{equation*}
and the projector on the right is just \((1 - \frac{\Delta - \frac{d}{2}}{m} n^{(R)}_{\mu'} \Gamma^{\mu'}(x')) \slashed{N}\) (in \eqref{eq:large-delta-simple-rindler-kernel} we instead just wrote the vector in the image of this projector), both in the limit \(z' \to 0\). These two matrices encode the vielbien choice, and the matrix on the right encodes the relationship between \(\Delta\) and mass. We can check that the kernel transforms properly between the two coordinate systems by using the spinor transformation corresponding to the change of vielbien from Poincare to Rindler coordinates, which is
\begin{equation*}
v_- =
\begin{pmatrix}
- \tanh \frac{t_R}{2} \\
e^{\rho}
\end{pmatrix}, \;
\Lambda_{1/2} = \frac{- \gamma^0 v^0_- + \gamma^1 v^1_-}{\sqrt{\eta_{ab} v_-^a v_-^b}} \gamma^2.
\end{equation*}

We also show that the form of the kernel is fixed where it is nonzero by demanding diffeomorphism invariance for the bulk spinor, while the boundary spinor is a quasi-primary operator of dimension \(\Delta\). To find equations for the kernel, we use that the unitaries generating bulk isometries generate boundary conformal transformations. The bulk field should transform according to the flow generated by the appropriate vector field. Concretely, we fix an orthonormal frame \(e_a\), and consider the transformation generated by the flow of a Killing vector \(\xi\). Under the pushforward by this flow, the components of \(e_a\) change by \(- \mathcal{L}_{\xi} e_a = - [\xi, e_a]\). Since \(\xi\) is a Killing vector, the generator \(J_{(\xi) a b} = \langle e_a, \mathcal{L}_{\xi} e_b \rangle\) is antisymmetric. A bulk field in a representation \(\rho\) of the spin group transforms by
\begin{equation*}
- \mathcal{L}^B_{\xi} \psi(x, z) = - (\rho(J_{(\xi) a b}) + \xi) \psi(x, z).
\end{equation*}
This is the flow generated by the operator \(\hat{\xi}\), and in the case of \(\ads_{d + 1}\), the same operator generates conformal transformations on the boundary. Then if the bulk field is written
\begin{equation*}
\psi(x, z) = \int d^d y K(x, z | y) \chi(y),
\end{equation*}
and the conformal transformation acts on \(\chi(y)\) by
\begin{equation*}
[\hat{\xi}, \chi(y)] = - \mathcal{L}_{\xi} \chi(y)
\end{equation*}
we find the operator constraints
\begin{equation}
\label{eq:generic-kernel-symmetry-constraint}
\mathcal{L}^{B(x,z)}_{\xi} K(x, z | y) = K(x, z | y) \circ \mathcal{L}^{(y)}_{\xi}
\end{equation}
where on the Lie derivatives we have indicated which variables the differential parts can act on. Actually, these are too restrictive in general dimension (specifically odd \(\ads\)), where the difference from \eqref{eq:generic-kernel-symmetry-constraint} can be by terms that integrate to zero against a boundary field, but in even \(\ads\) this stronger constraint can be satisfied.

The transformation of boundary primary fields is particularly simple in Poincare coordinates. We can derive the constraints on the Poincare kernel,
\begin{align*}
K(x, z | y) &= K(x - y, z) \\
(2 x_j (\Delta - d) - (x^2 + z^2) \partial_{x^j} + 2 x^k \Sigma_{jk} + 2 z \Sigma_{jz}) K(x, z) &= 0 \\
(x^j \partial_j + z \partial_z) K(x, z) &= (\Delta - d) K(x, z) \\
(x_j \partial_{x^k} - x_k \partial_{x^j}) K(x, z) &= - (\Sigma_{jk} K(x, z) - K(x, z) \Sigma^{\partial}_{jk})
\end{align*}
where \(\Sigma_{jk}\) is the generator of the Lorentz group in the desired bulk representation, \(\Sigma^{\partial}_{jk}\) is the generator of the Lorentz group in the boundary representation, and \(\Delta\) is the dimension of the boundary field. The last equation shows that \(K(0, z)\) is an intertwiner for representations of the boundary Lorentz group. These equations can be used to find the form of the kernel for arbitrary spin. In the spinor case (we assume the spinor transforms irreducibly on the boundary), a non-zero solution
\begin{equation*}
K(x, z) = \frac{\slashed{x} + z \gamma^z}{\sqrt{x^j x_j + z^2}}
\left( \frac{z^2 + x^j x_j}{z} \right)^{\Delta - d} \iota_1
\end{equation*}
where \(\iota_1\) maps the boundary spinor representation into the \(\gamma^z = 1\) eigenspace of the bulk spinor representation (its presence and image as a particular eigenspace of \(\gamma^z\) is mandated by the fact that \(K(0, z)\) is an intertwiner for an irreducible representation of a Lorentz subgroup, so its image is irreducible and hence \(\gamma^z\) is constant on the image, but the choice of sign of eigenspace is arbitrary). This is the unique non-zero solution, up to choice of scale and the sign of the \(\gamma^z\) eigenspace for the irreducible representation. It is also straightforward to show directly from the constraints that \(K\) must satisfy a Dirac equation,
\begin{equation*}
\slashed{\nabla} K(x, z) = \left(\Delta - \frac{d}{2}\right) K(x, z).
\end{equation*}
\section{Numerics}
\label{app:numerics}
Here we note some practical formulas for numerics with our divergent kernels. If we have some boundary quantity \(F^{\partial}(u)\) that is linear in boundary fields, we need to compute
\begin{align*}
F^B(t_R, \rho) &= \int du I(t_R - u, \rho) F^{\partial}(u) \Theta(\text{spacelike}) \\
&= 2 e^{- \rho} \tilde{F}^B(t_R, \rho) \\
\tilde{F}^B(t_R, \rho) &= \int_{-1}^1 \tilde{I}(x, \rho) F^{\partial}(t_R - u(x)) dx \\
\tilde{I}(x, \rho) &= \frac{1}{1 + e^{- \rho} x} \frac{1}{(1 - (e^{-\rho} x)^{2})^{\Delta}} (1 - x^2)^{\Delta - 1/2} \\
u_{\rho}(x) &= \ln \frac{1 + e^{-\rho} x}{1 - e^{-\rho} x}.
\end{align*}
We will also need the partial derivatives
\begin{align*}
\partial_{t_R} \tilde{F}^B(t_R, \rho) &= \int_{-1}^1 dx \tilde{I}(x, \rho) 
\partial_{t_R} F^{\partial}(t_R - u_{\rho}(x)) \\
\partial_{\rho} \tilde{F}^B(t_R, \rho) &=
\int_{-1}^1 dx \tilde{I}(x, \rho)
\frac{e^{- \rho} x}{1 - (e^{-\rho} x)^2}
\left(1 - (2 \Delta + 1) e^{-\rho} x + 2 \partial_{t_R} \right)
F^{\partial}(t_R - u_{\rho}(x)).
\end{align*}
Finally, to make the integral against \((1-x^2)^{\Delta - 1/2}\) manifestly convergent, we can use that
\begin{align*}
\int_{-1}^1 dx (1 - x^2)^{\Delta - 1/2} f(x)
&= \int_{-\pi/2}^{\pi/2} d \theta (\cos \theta)^{\frac{1}{2}+\Delta} f(\sin \theta) \\
&= \frac{1}{\frac{1}{2} + \Delta}
\int_0^1 \frac{dt}{(2 - t^{1/(1/2 + \Delta)})^{1/2 - \Delta}}
(f(1 - t^{1/(1/2 + \Delta)}) + f(t^{1/(1/2 + \Delta)} - 1)).
\end{align*}

The bulk size distribution is
\begin{equation*}
\mathcal{G}^B_{\mu}(t_R, \rho)_j = \int du du' K_{\Delta}(t_R - u, \rho)_j K_{\Delta}(t_R - u', \rho)_j \mathcal{G}^{\partial}_{\mu}(u, u').
\end{equation*}
We can symmetrize the integrand, which is the same as taking the real part in a basis where all the \(\gamma^j\) are real. Following the above, the bulk quantity can be written in terms of
\begin{equation*}
H_{\mu}(t_R, \rho, t_R', \rho')
= \int_{-1}^1 dx dx' \tilde{I}(x, \rho) \tilde{I}(x', \rho') \Re [\mathcal{G}^{\partial}_{\mu}(t_R - u_{\rho}(x), t_R' - u_{\rho'}(x'))]
\end{equation*}
and its derivatives. This function satisfies the identity \(H_{\mu}(t_R, \rho, t_R', \rho') = H_{\mu}(t_R', \rho', t_R, \rho)\). 
\subsection{Chebyshev polynomial method}
\label{sec:chebyshev-polynomial-numerics}
Another strategy to numerically integrate against the kernel is to expand in the complete Chebyshev polynomials \(T_n(x)\), and use the analytic continuation of
\begin{align*}
\int_{-1}^1 (1-x^2)^{\alpha} T_n(x) dx &= 2^{2 \alpha + 1} B(\alpha + 1, \alpha + 1)
\pFq{3}{2} \left( -n, n, \alpha + 1; \frac{1}{2}, 2 \alpha + 2; 1 \right) \\
&\xrightarrow[\alpha \to \Delta - 3/2]{}
\begin{cases}
0 & n \text{ odd} \\
(-1)^{\frac{n}{2}}
\frac{4^{1-\Delta} \pi}{2 \Delta - 1} 
\frac{1}{B(\Delta + \frac{n}{2}, \Delta - \frac{n}{2})} & n \text{ even}
\end{cases},
\end{align*}
where we can define the integral for $\alpha < 1$ by a figure-eight contour.

\end{document}